\documentclass[11pt,a4paper]{article}
\pdfoutput=1
\usepackage[utf8]{inputenc}
\usepackage{jheppub}
\hypersetup{unicode=true,bookmarksopen=true}

\usepackage{caption}
\usepackage{subcaption}

\usepackage{amsthm}
\usepackage{mathtools}
\usepackage{lmodern}
\usepackage[T1]{fontenc}
\usepackage{bbm}
\DeclareMathAlphabet{\mathbfi}{OML}{cmm}{b}{it}
\usepackage[scr=rsfso]{mathalpha}

\let\originalleft\left
\let\originalright\right
\renewcommand{\left}{\mathopen{}\mathclose\bgroup\originalleft}
\renewcommand{\right}{\aftergroup\egroup\originalright}
\makeatletter
\newcommand{\biggg}{\bBigg@\thr@@}
\newcommand{\Biggg}{\bBigg@{3.5}}
\makeatother

\makeatletter
\newenvironment{equations}[1][]{\subequations\ifx\relax#1\relax\else\label{#1}\fi\align\ignorespaces}{\endalign\ignorespacesafterend\endsubequations}
\def\@spliteq#1{\begin{equation}\begin{split}#1\end{split}\end{equation}}
\def\@spliteqstar#1{\begin{equation*}\begin{split}#1\end{split}\end{equation*}}
\def\splitequation{\collect@body\@spliteq}
\expandafter\def\csname splitequation*\endcsname{\collect@body\@spliteqstar}

\expandafter\def\csname endsplitequation*\endcsname{\ignorespacesafterend}
\makeatother

\renewcommand{\vec}[1]{{\ifnum9<1#1\mathbf{#1}\else\ifcat\noexpand#1\relax\boldsymbol{#1}\else\mathbfi{#1}\fi\fi}}
\newcommand{\mathe}{\mathrm{e}}
\newcommand{\mathi}{\mathrm{i}}
\let\oldre\Re
\let\oldim\Im
\renewcommand{\Re}{\oldre\mathfrak{e}\,}
\renewcommand{\Im}{\oldim\mathfrak{m}\,}
\newcommand{\total}{\mathop{}\!\mathrm{d}}

\newcommand{\abs}[1]{{\left\lvert{#1}\right\rvert}}
\newcommand{\norm}[1]{{\left\lVert{#1}\right\rVert}}
\newcommand{\sgn}{\operatorname{sgn}}

\newcommand{\arccot}{\operatorname{arccot}}

\newcommand{\eqend}[1]{\,#1}
\newcommand{\bigo}[1]{\mathcal{O}\left({#1}\right)}

\makeatletter
\def\bra{\@ifnextchar[\bra@size\bra@nosize}
\def\bra@size[#1]#2{#1\langle{#2}#1\rvert}
\def\bra@nosize#1{\left\langle{#1}\right\rvert}
\def\ket{\@ifnextchar[\ket@size\ket@nosize}
\def\ket@size[#1]#2{#1\lvert{#2}#1\rangle}
\def\ket@nosize#1{\left\lvert{#1}\right\rangle}
\makeatother
\newcommand{\expect}[1]{\left\langle{#1}\right\rangle}

\newcommand{\op}{\mathcal{O}}
\newcommand{\normord}[1]{\mathopen{:}{#1}\mathclose{:}}

\newcommand{\supp}{\operatorname{supp}}

\makeatletter
\DeclareRobustCommand*{\citeDLMFeq}{\hyper@normalise\citeDLMFeq@}
\def\citeDLMFeq@#1#2{\cite[\hyper@linkurl{Eq.~#1.#2}{https://dlmf.nist.gov/#1.E#2}]{DLMF}}
\makeatother

\allowdisplaybreaks

\makeatletter
\gdef\@fpheader{\strut}
\makeatother

\begin{document}

\title{The Sine--Gordon QFT in de~Sitter spacetime}

\author[1]{Daniela Cadamuro}
\author[1]{\!\!,\ Markus B. Fr{\"o}b}
\author[2]{and Carolina Moreira Ferrera}

\affiliation[1]{Institut f{\"u}r Theoretische Physik, Universit{\"a}t Leipzig, Br{\"u}derstra{\ss}e 16, 04103 Leipzig, Germany}

\affiliation[2]{School of Physics, HH Wills Laboratory, University of Bristol, Tyndall Avenue, Bristol BS8 1TL, United Kingdom}

\emailAdd{cadamuro@itp.uni-leipzig.de}
\emailAdd{mfroeb@itp.uni-leipzig.de}
\emailAdd{carolina.moreiraferrera@bristol.ac.uk}

\abstract{We consider the massless Sine--Gordon model in de Sitter spacetime, in the regime $\beta^2 < 4 \pi$ and using the framework of perturbative algebraic quantum field theory. We show that a Fock space representation exists for the free massless field, but that the natural one-parameter family of vacuum-like states breaks the de Sitter boost symmetries. We prove convergence of the perturbative series for the S matrix in this representation, and construct the interacting Haag--Kastler net of local algebras from the relative S matrices. We show that the net fulfills isotony, locality and de Sitter covariance (in the algebraic adiabatic limit), even though the states that we consider are not invariant. We furthermore prove convergence of the perturbative series for the interacting field and the vertex operators, and verify that the interacting equation of motion holds.}


\maketitle

\section{Introduction}

The massless Sine--Gordon model is a two-dimensional interacting quantum field theory. Its classical action (in a curved spacetime) is given by
\begin{equation}
\label{eq:intro_sinegordon_action}
S = \int \left[ - \frac{1}{2} \nabla^\mu \phi \nabla_\mu \phi + 2 g \cos(\beta \phi) \right] \sqrt{-g} \total^2 x \eqend{,}
\end{equation}
where $\beta > 0$ is the coupling constant and $g$ is the interaction cutoff. The quantization of this model has been treated in various approaches, including the framework of Euclidean Constructive Quantum Field Theory, where convergence of Euclidean correlation functions of certain interacting fields has been shown for various ranges of the coupling constant both in finite and infinite volume, see references in~\cite{froebcadamuro2022a}. Since the classical equation of motion
\begin{equation}
\nabla^2 \phi - 2 \beta g \sin(\beta \phi) = 0
\end{equation}
admits an infinite number of conserved charges in flat spacetime~\cite{zakharovtakhtadzhyanfaddeev1974}, the classical model is integrable and one expects that integrability also holds in the quantum theory, namely that there exists an infinite number of conserved currents~\cite{flume1975,yoon1976,zanello2023}. The corresponding S matrix has been conjectured in the form factor programme~\cite{zamolodchikov1979,babujianetal1999,babujiankarowski2002}. In this approach, Wightman $n$-point functions of interacting pointlike local fields are computed given the input of the two-particle S matrix, which for an integrable model characterizes the particle interaction completely. On the other hand, in the Algebraic Quantum Field Theory (AQFT) framework, where a model is characterized in terms of the $C^\star$-algebras of local observables obeying the Haag--Kastler axioms, first steps towards the construction of the massless Sine-Gordon model were obtained by using observables localized in infinite extended regions called wedges~\cite{cadamurotanimoto2018}. In this approach the passage to strictly local observables is obtained by abstract arguments of Tomita--Takesaki modular theory, though this step is not yet under control for the Sine--Gordon model. The conjectured S matrix of the massless Sine--Gordon model has a very rich particle spectrum, including solitons, anti-solitons, and their finitely many bound states called breathers. However, in the rigorous AQFT constructions there is no proof of the factorization of the S matrix into products of two-particle S matrices, compare Ref.~\cite{bahnsrejzner2018} where one finds a proof of the convergence of the perturbative S matrix with IR cutoff, but its factorization has not been proven. The Sine--Gordon model is also conjectured to be equivalent to the massive Thirring model (Coleman's equivalence~\cite{coleman1975}), where the solitons and anti-solitons in the massless Sine-Gordon model seem to be related to certain fermionic solutions in the massive Thirring model. This equivalence has been proven in the massive and massless cases for $\beta^2 < 4\pi$ and in finite volume~\cite{froehlichseiler1976,benfattofalcomastropietro2009}, and for $\beta^2 = 4 \pi$ in finite~\cite{dimock1998} and infinite volume~\cite{bauerschmidtwebb2020}. The massless Sine-Gordon model has also been treated in the context of stochastic quantisation and its relation to the Euclidean correlation functions in the quantum theory has been investigated~\cite{hairershen2014,chandrahairershen2018,ohrobertsosoewang2021,barashkov2022}.

Here, we focus on the perturbative AQFT framework, which is reviewed for example in Ref.~\cite{fredenhagenrejzner2016}. In particular, we use the Bogoliubov formula to define interacting field operators. This approach combines methods from perturbation theory and the framework of AQFT, which means in particular that the construction of the $\ast$-algebra of local observables is achieved using perturbative methods. The main challenges of this approach are to show convergence of the formal perturbation series of the local observables, and to choose a state to represent the observable algebra on a Hilbert space via the analogue of the GNS construction for $\ast$-algebras, see for example~\cite[Thm.~1]{khavkinemoretti2015}. In the case of the massless Sine--Gordon model results have been achieved in both these directions. Indeed, in Ref.~\cite{bahnsrejzner2018} the convergence of the perturbation series of the $S$-matrix with fixed interaction cutoff, which is given as a formal power series in $g$ (or $\hbar$), has been shown in the regime $\beta^2 < 4\pi$. The same proof of convergence has been achieved for the derivative of the interacting field $\partial_\mu \phi$ and of the vertex operators $V_{\pm \beta} = \mathe^{\pm \mathi \beta \phi}$. Since the vacuum state of the free scalar field in two dimensions is affected by infrared (IR) problems as the field becomes massless, these results were obtained with a fixed IR cutoff, where the positivity of the state is only guaranteed when smearing with test functions with vanishing mean. In later work~\cite{bahnsfredenhagenrejzner2021}, a new state (the Derezi{\'n}ski--Meissner state) has been introduced, which is a proper positive Hadamard state without IR cutoff for the free massless scalar field, and its equivalence with the Schubert construction has been shown. In this representation of the observable algebra, the authors could then show that the S matrix is unitary and satisfies Bogoliubov's causal factorization condition. Using this result, they constructed a family of unitary operators, namely the relative S-matrices, which generate the local algebras of observables of the model. Lastly, the equivalence of the Sine--Gordon theory with the massive Thirring model (Coleman's equivalence) was discussed in the pAQFT framework.

In more recent work~\cite{froebcadamuro2022a,froebcadamuro2022b}, we showed in the pAQFT framework the convergence of the renormalized expectation values of $\partial_\mu \phi \partial_\nu \phi$ and of the stress-energy tensor $T_{\mu\nu}$, in a large class of quasi-free Hadamard states regularized with IR and UV cutoffs. We showed that convergence holds after removal of the IR and UV cutoffs but without removal of the adiabatic interaction cutoff $g$, and also in the regime $\beta^2 < 4 \pi$.

The goal of the present work is to extend the proof of convergence of the S matrix, of the vertex operators and of the interacting field, as well as the contruction of the local algebras, to a curved spacetime, namely the two-dimensional de Sitter spacetime. To the best of our knowledge, our result represents one of the few interacting quantum field theories established on a curved spacetime after the $P(\phi)_2$-model on de Sitter spacetime constructed by Figari, H{\"o}egh-Krohn and Nappi~\cite{figarihoeghkrohnnappi1975}, which was later supplemented with new non-perturbative results and mathematical insights by Barata, J{\"a}kel and Mund~\cite{baratajaekelmund2019}. (Four-dimensional) de Sitter spacetime is important as a model for both the primordial and the present accelerated expansion of our universe. Moreover, it possesses the maximum number of symmetries (three in two dimensions), and is thus one of the simplest curved spacetimes. For this reason, Ref.~\cite{chakrabortystout2023} recently studied the leading perturbative correction coming from the Sine--Gordon interaction in de~Sitter spacetime, but our work represents the first full construction of this model.

Using the pAQFT framework, we construct the Sine-Gordon model as a perturbation of the free massless scalar field. As compared to the construction in Minkowski spacetime, the construction in de Sitter spacetime has the following differences: Since on de Sitter spatial sections are compact, there is no infrared divergence in the vacuum two-point function of the massless scalar field, and one can work with the field itself rather than its derivative. This also allows us to use the ordinary Fock space representation of the massless field in terms of creation and annihilation operators, rather than the Derezi{\'n}ski--Meissner representation or an analogue of it. However, even though we consider a one-parameter family of vacuum-like states (in the sense that they are annihilated by all the annihilation operators in the mode expansion of the massless field), it turns out that all these states break de Sitter invariance.

The remainder of this work is organized as follows: we introduce the mode expansion of the massless field and the family of states that we consider in Sec.~\ref{sec:desitter}. In Sec.~3 we show convergence of the Bogoliubov S matrix strongly on a dense domain of vectors generated from the above family, without removal of the interaction cutoff, and show that it is unitary on this domain. Starting from this, we construct the relative S matrix fulfilling the property of causal factorization, and we use this to generate an Haag-Kastler net satisfying the properties of isotony, locality and covariance. This is possible in spite of the interaction cutoff by taking the so called algebraic adiabatic limit, namely by considering the equivalence classes of relative S-matrices where the interaction cutoff (i.e., the smearing function) is constant in the localization region. In other words, though the family of states are not invariant under the de Sitter symmetries, the net still retains the covariance property. Finally, we show convergence of the perturbative series of the interacting field given by the Bogoliubov formula on the same dense domain as above.

\section{Free massless scalar field on de~Sitter}
\label{sec:desitter}

\subsection{The de~Sitter spacetime}

The $n$-dimensional de~Sitter spacetime can be embedded in an $(n+1)$-dimensional Minkowski spacetime with Cartesian coordinates $X^A$ as the hyperboloid
\begin{equation}
\label{eq:desitter_hyperboloid}
\eta_{AB} X^A X^B = H^{-2} \eqend{,}
\end{equation}
where $H$ is the Hubble rate or inverse de~Sitter radius. In this form, it is easy to see that the Lorentz symmetries of the embedding Minkowski become symmetries of de~Sitter when restricted to the hyperboloid. Namely, the generators $M^{AB} = X^A \partial_{X_B} - X^B \partial_{X_A}$ are tangent to the hyperboloid, since they leave the condition~\eqref{eq:desitter_hyperboloid} invariant. It thus follows that de~Sitter is a maximally symmetric spacetime with $n(n+1)/2$ Killing vectors, which are the restrictions of $M_{AB}$ to the hyperboloid.

There exist various intrinsic coordinate systems which cover all or part of the full de~Sitter spacetime. For our purposes, it is useful to work in coordinates that cover the whole manifold, and which make manifest the fact that de~Sitter is conformally flat. Restricting to $n = 2$ dimensions, these are obtained by taking
\begin{equation}
\label{eq:desitter_embedding}
X^0 = - \frac{\cot(\tau)}{H} \eqend{,} \quad X^1 = \frac{\cos(\theta)}{H \sin(\tau)} \eqend{,} \quad X^2 = \frac{\sin(\theta)}{H \sin(\tau)}
\end{equation}
with $\theta \in [0,2\pi)$ and $\tau \in (0,\pi)$. Clearly they fulfill the condition~\eqref{eq:desitter_hyperboloid}, and the induced de~Sitter metric $g_{\mu\nu}$ is obtained from
\begin{equation}
\total s^2 = \eta_{AB} \total X^A \total X^B = \frac{- \total \tau^2 + \total \theta^2}{H^2 \sin^2(\tau)} = g_{\mu\nu} \total x^\mu \total x^\nu \eqend{,}
\end{equation}
where $x^\mu = (\tau,\theta)$ are the intrinsic coordinates of de~Sitter spacetime.

The rotational symmetry $\theta \to \theta + a$ is generated by the Killing vector
\begin{equation}
\label{eq:desitter_killing_rot}
\xi_\text{rot} \equiv \xi_\text{rot}^\mu \partial_\mu = \partial_\theta \eqend{,}
\end{equation}
which is the restriction of $M^{12}$ to the hyperboloid. To compute the restriction, we invert the embedding~\eqref{eq:desitter_embedding} to obtain
\begin{equation}
\label{eq:desitter_embedding_inversion}
\tau = \arccot(- H X^0) \eqend{,} \quad \theta = \arctan\left( \frac{X^2}{X^1} \right) \eqend{,}
\end{equation}
and then use the chain rule. While the inversion~\eqref{eq:desitter_embedding_inversion} itself is not unique, evaluating the result on the hyperboloid gives a unique and well-defined tangent vector in de~Sitter. The other two (boost) symmetries are generated by $M^{01}$ and $M^{02}$, whose restriction gives the Killing vectors
\begin{equations}[eq:desitter_killing_boost]
\xi_\text{boost,1} &\equiv \xi_\text{boost,1}^\mu \partial_\mu = \sin(\tau) \cos(\theta) \partial_\tau + \cos(\tau) \sin(\theta) \partial_\theta \eqend{,} \\
\xi_\text{boost,2} &\equiv \xi_\text{boost,2}^\mu \partial_\mu = \sin(\tau) \sin(\theta) \partial_\tau - \cos(\tau) \cos(\theta) \partial_\theta \eqend{.}
\end{equations}
Unlike for rotations, the finite transformations for boosts have a complicated expression, which we derive in appendix~\ref{sec:appendix_ds}. We obtain
\begin{equation}
\label{eq:desitter_finite_transformation}
\mathe^{a \xi_\text{rot} + b \xi_\text{boost,1} + c \xi_\text{boost,2}} f(\tau,\theta) = f(\tau_{abc}, \theta_{abc})
\end{equation}
with $\tau_{abc}$ and $\theta_{abc}$ given by~\eqref{eq:appendix_ds_coordinates}.

Finally, a straightforward calculation shows that the Killing vectors they form a representation of the de~Sitter symmetry algebra $\mathfrak{so}(2,1)$:
\begin{equation}
\label{eq:desitter_killing_algebra}
[ \xi_\text{rot}, \xi_\text{boost,1} ] = - \xi_\text{boost,2} \eqend{,} \quad [ \xi_\text{rot}, \xi_\text{boost,2} ] = \xi_\text{boost,1} \eqend{,} \quad [ \xi_\text{boost,1}, \xi_\text{boost,2} ] = \xi_\text{rot} \eqend{.}
\end{equation}

\subsection{Free massless scalar field}

On the de~Sitter spacetime, we now consider a free massless, minimally coupled scalar field $\phi$ with action
\begin{equation}
\label{eq:free_scalar_action}
S = - \frac{1}{2} \int \nabla^\mu \phi \nabla_\mu \phi \sqrt{-g} \total^2 x = \frac{1}{2} \int \left[ ( \partial_\tau \phi )^2 - ( \partial_\theta \phi )^2 \right] \total \tau \total \theta \eqend{.}
\end{equation}
We see that the action is the same as for a massless scalar field in flat space, i.e., $\phi$ does not see the curvature of the de~Sitter spacetime. This arises from the fact that in two dimensions minimal coupling to the Ricci scalar is equivalent to conformal coupling, and hence the action is (classically) conformally invariant.

The equation of motion that follows from the action~\eqref{eq:free_scalar_action} reads
\begin{equation}
\label{eq:free_scalar_eom}
\partial^2 \phi \equiv \left( - \partial_\tau^2 + \partial_\theta^2 \right) \phi = 0 \eqend{,}
\end{equation}
and one verifies its invariance under the de~Sitter symmetries: we have $[ \xi_\text{rot}, \partial^2 ] = 0$, $[ \xi_\text{boost,1}, \partial^2 ] = - 2 \cos(\tau) \cos(\theta) \partial^2$ and $[ \xi_\text{boost,2}, \partial^2 ] = - 2 \cos(\tau) \sin(\theta) \partial^2$, such that solutions are mapped into solutions by the action of the Killing vectors. We can thus decompose the solution into eigenmodes of the various Killing vectors. However, since they do not commute~\eqref{eq:desitter_killing_algebra}, these decompositions are not orthogonal, and we have to choose a maximal commuting set to obtain an orthogonal decomposition. The simplest such set is given by the single Killing vector $\xi_\text{rot}$, whose eigenmodes and eigenvalues depend on the boundary conditions that one imposes on the field $\phi$. Since two-dimensional de~Sitter spacetime is not simply connected, it is possible to impose non-trivial boundary conditions which lead to interesting effects~\cite{epsteinmoschella2021,higuchischmiedingblanco2023}. Here we focus on the simplest case, namely periodic boundary conditions $\phi(\tau,\theta+2\pi) = \phi(\tau,\theta)$, such that the eigenmodes of $\xi_\text{rot}$ are proportional to $\mathe^{\mathi n \theta}$ with $n \in \mathbb{Z}$.

The solutions of the equation of motion~\eqref{eq:free_scalar_eom} corresponding to the eigenvalue $\mathi n$ of $\xi_\text{rot}$ are then given by
\begin{equation}
\label{eq:free_scalar_fockmode}
\phi_n(\tau,\theta) = \left[ c_1(n) \mathe^{\mathi n \tau} + c_2(n) \mathe^{- \mathi n \tau} \right] \mathe^{\mathi n \theta}
\end{equation}
for $n \neq 0$, where the $c_i$ are constants. However, for $n = 0$ we have the solution
\begin{equation}
\label{eq:free_scalar_zeromode}
\phi_0(\tau,\theta) = c_3 + c_4 \tau \eqend{,}
\end{equation}
the so-called zero mode. In contrast to the flat Minkowski spacetime, where the massless scalar field has an infrared divergence and only derivatives of the field exist, no divergence exists for global de~Sitter spacetime (which has compact spatial sections), and the field itself is well-defined. However, we will see later that the zero mode~\eqref{eq:free_scalar_zeromode} is responsible for the breaking of de~Sitter invariance in the two-point function.

An inner product on a hypersurface of constant $\tau$ is given by
\begin{equation}
\expect{ f, g } \equiv \mathi \int \Big[ f^*(\tau,\theta) \partial_\tau g(\tau,\theta) - \partial_\tau f^*(\tau,\theta) g(\tau,\theta) \Big] \total \theta = - \expect{ g^*, f^* } \eqend{,}
\end{equation}
and is independent of $\tau$ if $f$ and $g$ are solutions of the equation of motion~\eqref{eq:free_scalar_eom}. We easily compute that modes for different $n$ are orthogonal:
\begin{equations}[eq:free_scalar_modeproduct]
\expect{ \phi_m, \phi_n } &= 4 \pi n \, \delta_{m,n} \left[ \abs{ c_2(n) }^2 - \abs{ c_1(n) }^2 \right] \eqend{,} \label{eq:free_scalar_modeproduct_mn} \\
\expect{ \phi_0, \phi_n } &= 0 \eqend{,} \\
\expect{ \phi_0, \phi_0 } &= 2 \pi \mathi \left( c_3^* c_4 - c_3 c_4^* \right) \eqend{,} \label{eq:free_scalar_modeproduct_00}
\end{equations}
where we used that $\int \mathe^{\mathi k \theta} \total \theta = 2 \pi \delta_{k,0}$.

\subsection{A one-parameter family of states}

Given the solutions~\eqref{eq:free_scalar_fockmode} and~\eqref{eq:free_scalar_zeromode} for the mode functions and their inner products~\eqref{eq:free_scalar_modeproduct}, we can now proceed to quantize the scalar field. The quantized field $\hat{\phi}$ is expanded in terms of creation and annihilation operators $\hat{a}_n^\dagger$ and $\hat{a}_n$ according to
\begin{equation}
\label{eq:free_scalar_fieldexpansion}
\hat{\phi}(\tau,\theta) = \sum_n \left[ \hat{a}_n \phi_n(\tau,\theta) + \hat{a}_n^\dagger \phi_n^*(\tau,\theta) \right] \eqend{,}
\end{equation}
and the ``vacuum'' state $\ket{0}$ corresponding to a specific choice of the mode functions is the one satisfying $\hat{a}_n \ket{0} = 0$ for all $n$. The usual positive-frequency mode functions are obtained by choosing $c_1(n) = \Theta(-n)/\sqrt{4 \pi \abs{n}}$ and $c_2(n) = \Theta(n)/\sqrt{4 \pi \abs{n}}$; with this choice, the inner product~\eqref{eq:free_scalar_modeproduct_mn} is positive and normalized to $1$.

For the zero mode, however, there is no canonical choice. Zero modes appear also for massless, minimally coupled scalar fields in higher-dimensional de~Sitter spacetime~\cite{allenfolacci1987,kirstengarriga1993}, and there exists a family of solutions depending on a parameter $\alpha > 0$ that preserves covariance under time reversals. This family is given by the choice $c_3 = \alpha/2 + \mathi/(4\alpha)$, $c_4 = - \mathi/(2 \pi \alpha)$, and then also the inner product~\eqref{eq:free_scalar_modeproduct_00} is positive and normalized to $1$. Let us denote the corresponding state by $\ket{0_\alpha}$.

Since the set of mode functions is complete and orthonormal with these choices, the usual commutation relations
\begin{equation}
\label{eq:canonical_commutator}
[ \hat{a}_n, \hat{a}_m^\dagger ] = \delta_{m,n} \eqend{,} \quad [ \hat{a}_n, \hat{a}_m ] = 0 = [ \hat{a}_n^\dagger, \hat{a}_m^\dagger ]
\end{equation}
give the canonical equal-time ones
\begin{equation}
[ \hat{\phi}(\tau,\theta), \hat{\pi}(\tau,\theta') ] = \mathi \delta(\theta-\theta')
\end{equation}
for the (quantized) field and its conjugate momentum $\pi \equiv \partial_\tau \phi$. For the two-point function, we compute
\begin{splitequation}
\label{eq:free_scalar_2pf}
\mathi G^+_\alpha(\tau,\theta,\tau',\theta') &\equiv \bra{0_\alpha} \hat{\phi}(\tau,\theta) \hat{\phi}(\tau',\theta') \ket{0_\alpha} = \sum_n \phi_n(\tau,\theta) \phi_n^*(\tau',\theta') \\
&= \frac{1}{4 \pi^2 \alpha^2} \left[ \pi \alpha^2 - \mathi \left( \tau - \frac{\pi}{2} \right) \right] \left[ \pi \alpha^2 + \mathi \left( \tau' - \frac{\pi}{2} \right) \right] \\
&\quad+ \lim_{\epsilon \to 0^+} \sum_{n=1}^\infty \frac{\mathe^{- \mathi n (\tau-\tau') - n \epsilon}}{2 \pi n} \cos[ n (\theta-\theta') ] \\
&= \frac{\alpha^2}{4} - \frac{\mathi}{4 \pi} \left( \tau - \tau' \right) + \frac{1}{4 \pi^2 \alpha^2} \left( \tau - \frac{\pi}{2} \right) \left( \tau' - \frac{\pi}{2} \right) \\
&\quad- \frac{1}{4 \pi} \lim_{\epsilon \to 0^+} \left[ \ln\left( 1 - \mathe^{- \mathi (\tau-\tau'+\theta-\theta'-\mathi \epsilon)} \right) + \ln\left( 1 - \mathe^{- \mathi (\tau-\tau'-\theta+\theta'-\mathi \epsilon)} \right) \right] \eqend{,}
\end{splitequation}
where we inserted a convergence factor $\mathe^{- n \epsilon}$ in the sum over $n$. We can now verify covariance under time reversals, which are the transformations $\tau \to \pi - \tau$~\cite{allenfolacci1987,kirstengarriga1993}. Since time reversal is implemented by an anti-unitary operator, we require $\mathi G^+_\alpha(\pi-\tau,\theta,\pi-\tau',\theta') = \left[ \mathi G^+_\alpha(\tau,\theta,\tau',\theta') \right]^*$, which is seen to hold. Moreover, we compute
\begin{splitequation}
\label{eq:free_scalar_2pf_log}
&\lim_{\epsilon \to 0^+} \left[ \ln\left( 1 - \mathe^{- \mathi (\tau-\tau'+\theta-\theta'-\mathi \epsilon)} \right) + \ln\left( 1 - \mathe^{- \mathi (\tau-\tau'-\theta+\theta'-\mathi \epsilon)} \right) \right] + \mathi \left( \tau - \tau' \right) \\
&\quad= \lim_{\epsilon \to 0^+} \ln\left[ 2 \cos(\tau-\tau'-\mathi \epsilon) - 2 \cos(\theta-\theta') \right] \\
&\quad= \ln\abs{ 2 \cos(\tau-\tau') - 2 \cos(\theta-\theta') } + \mathi \pi \Theta\big[ \cos(\theta-\theta') - \cos(\tau-\tau') \big] \sgn \sin(\tau-\tau') \eqend{,} \raisetag{4em}
\end{splitequation}
where we used the distributional identity (valid for $z \in \mathbb{R}$)
\begin{equation}
\label{eq:log_distribution}
\lim_{\epsilon \to 0^+} \ln(z + \mathi a \epsilon) = \ln\abs{z} + \mathi \pi \sgn(a) \Theta(-z) \eqend{.}
\end{equation}

While the state $\ket{0_\alpha}$ is invariant under rotations (essentially by construction), such that $\left[ \xi_\text{rot}(\tau,\theta) + \xi_\text{rot}(\tau',\theta') \right] G^+_\alpha(\tau,\theta,\tau',\theta') = 0$, it is not invariant under boosts. We compute explicitly
\begin{splitequation}
\label{eq:free_scalar_2pf_boost}
\left[ \xi_\text{boost,1}(\tau,\theta) + \xi_\text{boost,1}(\tau',\theta') \right] G^+_\alpha(\tau,\theta,\tau',\theta') &= \mathi \frac{2 \pi \alpha^2 \cos(\tau) + ( \pi - 2 \tau' ) \sin(\tau)}{8 \pi^2 \alpha^2} \cos(\theta) \\
&\quad+ \mathi \frac{2 \pi \alpha^2 \cos(\tau') + ( \pi - 2 \tau ) \sin(\tau')}{8 \pi^2 \alpha^2} \cos(\theta') \eqend{,}
\end{splitequation}
and a similar expression for $\xi_\text{boost,2}$. Even though the result simplifies in the limit $\alpha \to \infty$, it does not vanish. On the other hand, expectation values of derivatives of the field are covariant in the limit $\alpha \to \infty$, and we have explicitly
\begin{splitequation}
\label{eq:free_scalar_2pf_derivative_boost}
&\left[ \mathscr{L}_{\xi_\text{boost,1}(\tau,\theta)} + \mathscr{L}_{\xi_\text{boost,1}(\tau',\theta')} \right] \partial_\mu \partial_\nu' G^+_\alpha(\tau,\theta,\tau',\theta') \\
&\qquad= \frac{\mathi}{4 \pi^2 \alpha^2} \begin{pmatrix} - 2 \cos(\tau) \cos(\theta) & \sin(\tau') \sin(\theta') \\ \sin(\tau) \sin(\theta) & 0 \end{pmatrix}_{\mu\nu} \to 0 \quad (\alpha \to \infty) \eqend{,}
\end{splitequation}
where $\mathscr{L}_\xi$ denotes the Lie derivative with respect to the vector field $\xi$, and a similar expression for $\xi_\text{boost,2}$.

The time-ordered two-point function or Feynman propagator $G^\text{F}_\alpha$ is obtained from the two-point function~\eqref{eq:free_scalar_2pf} according to
\begin{splitequation}
\label{eq:free_scalar_feynman}
\mathi G^\text{F}_\alpha(\tau,\theta,\tau',\theta') &\equiv \Theta(\tau-\tau') \mathi G^+_\alpha(\tau,\theta,\tau',\theta') + \Theta(\tau'-\tau) \mathi G^+_\alpha(\tau',\theta',\tau,\theta) \\
&= \frac{\alpha^2}{4} - \frac{\mathi}{4 \pi} \abs{\tau-\tau'} + \frac{1}{4 \pi^2 \alpha^2} \left( \tau - \frac{\pi}{2} \right) \left( \tau' - \frac{\pi}{2} \right) \\
&\quad- \frac{1}{4 \pi} \lim_{\epsilon \to 0^+} \left[ \ln\left( 1 - \mathe^{- \mathi (\abs{\tau-\tau'}+\theta-\theta'-\mathi \epsilon)} \right) + \ln\left( 1 - \mathe^{- \mathi (\abs{\tau-\tau'}-\theta+\theta'-\mathi \epsilon)} \right) \right] \eqend{.} \raisetag{4.4em}
\end{splitequation}
and we see that time-ordering amounts to replacing $\tau-\tau' \to \abs{\tau-\tau'}$. Analogously, the anti-time-ordered two-point function or Dyson propagator $G^\text{D}_\alpha$ is obtained by replacing $\tau-\tau' \to - \abs{\tau-\tau'}$.

The choice of the positive-frequency modes for all $n$ ensures that our family of states satisfies the Hadamard condition, which is necessary to obtain finite correlation functions for all operators (including composite ones such as the stress tensor)~\cite{fewsterverch2013}. A quasi-free state satisfies the Hadamard condition if a certain spectral condition on the wave front set of the two-point function holds, which is equivalent to the Hadamard form of the two-point function if the two points lie in a convex normal neighbourhood~\cite{radzikowski1996,moretti2021}. The latter condition is often easier to verify in practice, and in two dimensions it means that the Feynman propagator has the form~\cite{decaninifolacci2008}
\begin{equation}
\label{eq:free_scalar_hadamard}
\mathi G^\text{F}(x,x') = \frac{1}{4 \pi} \lim_{\epsilon \to 0^+} \Bigl[ V(x,x') \ln\bigl[ M \sigma(x,x') + \mathi \epsilon \bigr] + W(x,x') \Bigr] \eqend{,}
\end{equation}
where $V$ and $W$ are smooth symmetric~\cite{moretti2000,hackmoretti2012,kaminski2019} biscalars, $\sigma(x,x')$ is the Synge world function equal to one half of the square of the geodesic distance $\mu(x,x')$ between the points $x$ and $x'$, and $M$ is an arbitrary scale to make the argument of the logarithm dimensionless. The biscalar $W$ incorporates the state dependence and is quite arbitrary, but $V$ is uniquely determined by the geometry and thus in the maximally symmetric de Sitter spacetime a function of $\mu$. To determine it, we use that
\begin{equation}
\label{eq:free_scalar_mu_relation}
\nabla^\rho \mu \nabla_\rho \mu = 1 \eqend{,} \quad \nabla^2 \mu = H \cot(H \mu)
\end{equation}
holds in two-dimensional de Sitter spacetime for the geodesic distance $\mu = \mu(x,x')$ itself~\cite{allenjacobson1986}. For a massless scalar field, $V$ satisfies $\nabla^2 V = 0$ and $\lim_{x' \to x} V(x,x') = - 1$~\cite{decaninifolacci2008}, which has the unique solution $V = - 1$. To compare the form~\eqref{eq:free_scalar_hadamard} with our Feynman propagator~\eqref{eq:free_scalar_feynman}, we need to express $\mu$ as a function of the coordinates $x$ and $x'$. Defining $z \equiv \cos(H \mu)$, the first relation in~\eqref{eq:free_scalar_mu_relation} results in $\nabla_\rho z \nabla^\rho z = H^2 (1-z^2)$, and a long but straightforward computation verifies that
\begin{equation}
z(x,x') = 1 + \frac{\cos(\theta - \theta') - \cos(\tau - \tau')}{\sin(\tau) \sin(\tau')} = z(x',x)
\end{equation}
is the unique function that satisfies this relation. Using then the relation~\eqref{eq:free_scalar_2pf_log} and the definition~\eqref{eq:free_scalar_feynman} of the Feynman propagator in terms of the two-point functions, it follows that
\begin{splitequation}
\mathi G^\text{F}_\alpha(\tau,\theta,\tau',\theta') &= \frac{\alpha^2}{4} + \frac{1}{4 \pi^2 \alpha^2} \left( \tau - \frac{\pi}{2} \right) \left( \tau' - \frac{\pi}{2} \right) \\
&\quad- \frac{1}{4 \pi} \lim_{\epsilon \to 0^+} \ln\bigl[ 2 (1-z) \sin(\tau) \sin(\tau') + \mathi \epsilon \bigr] \eqend{,}
\end{splitequation}
where we used that $\sin(\tau) \geq 0$ since $\tau \in (0,\pi)$. Comparing with the form~\eqref{eq:free_scalar_hadamard}, we find
\begin{equation}
W(x,x') = \pi \alpha^2 + \frac{1}{\pi \alpha^2} \left( \tau - \frac{\pi}{2} \right) \left( \tau' - \frac{\pi}{2} \right) - \ln\left[ 4 \sin(\tau) \sin(\tau') \right] - \ln\left[ \frac{1 - \cos(H \mu)}{M \mu^2} \right] \eqend{,}
\end{equation}
which is a smooth function of the coordinates, and thus the one-parameter family of states that we have constructed satisfies the Hadamard condition.

\subsection{Minkowski limit}

For this limit, we have to pass to new coordinates $t$ and $x$ according to~\cite{candelasdowker1979,froeb2023}
\begin{equations}
\tau &= \pi + \arctan\left( H x - \mathe^{- H t} \right) - \arctan\left( H x + \mathe^{- H t} \right) \eqend{,} \\
\theta &= \pi - \arctan\left( H x - \mathe^{- H t} \right) - \arctan\left( H x + \mathe^{- H t} \right) \eqend{,}
\end{equations}
in which the de~Sitter metric $g_{\mu\nu}$ reads
\begin{equation}
\total s^2 = g_{\mu\nu} \total x^\mu \total x^\nu = - \total t^2 + \mathe^{2 H t} \total x^2 \eqend{.}
\end{equation}
With $t,x \in \mathbb{R}$, these coordinates cover half of the hyperboloid, namely the so-called expanding Poincar{\'e} patch. There, the Minkowski limit can be easily taken by sending $H \to 0$. Transforming the two-point function~\eqref{eq:free_scalar_2pf} into this coordinate system, we obtain
\begin{splitequation}
\label{eq:free_scalar_2pf_minkowski}
\mathi G^+_\alpha(t,x,t',x') &= \frac{\alpha^2}{4} - \frac{1}{4 \pi} \ln\left[ \mathi H ( t-t' + x-x' - \mathi \epsilon ) \right] \\
&\quad- \frac{1}{4 \pi} \ln\left[ \mathi H ( t-t' - x+x' - \mathi \epsilon ) \right] + \bigo{H} \eqend{,}
\end{splitequation}
which for a suitable choice of $\alpha$ agrees with the flat-space two-point function in the small-mass limit~\cite[Eq.~(113)]{froebcadamuro2022a}. Namely, we need to take
\begin{equation}
\alpha = \sqrt{ \frac{2}{\pi} \ln\left( \frac{m \mathe^\gamma}{2 H} \right) } \eqend{,}
\end{equation}
and then the flat-space limit $H \to 0$ yields $\alpha \to \infty$. Here, $m$ is the mass of the scalar field in Minkowski space, and we recover the well-known logarithmic infrared divergence of the free scalar field in flat space as $m \to 0$. We see that the parameter $\alpha$ plays in de~Sitter spacetime a role similar to the mass $m$ in Minkowski spacetime: for any finite $\alpha$ or any finite mass $m$, the two-point function and thus the state is positive as required, but the limit $\alpha \to \infty$ (or $m \to 0$) is singular. Moreover, if one only considers the leading terms in these limits and neglects the terms that would vanish, the two-point function ceases to be positive.

Furthermore, we can also consider the coordinate transformation
\begin{equations}
\tau &= 2 \arctan \mathe^{H T} \eqend{,} \\
\theta &= \frac{\pi}{2} + \arctan(H X) \eqend{,}
\end{equations}
and in these coordinates the de~Sitter metric $g_{\mu\nu}$ reads
\begin{equation}
\total s^2 = g_{\mu\nu} \total x^\mu \total x^\nu = - \total T^2 + \frac{\cosh^2(H T)}{(1+H^2 X^2)^2} \total X^2 \eqend{.}
\end{equation}
With $T,X \in \mathbb{R}$ the full hyperboloid is now covered, and again the Minkowski limit can be easily taken by sending $H \to 0$. Transforming the two-point function~\eqref{eq:free_scalar_2pf} into this coordinate system, we obtain
\begin{splitequation}
\mathi G^+_\alpha(t,x,t',x') &= \frac{\alpha^2}{4} - \frac{1}{4 \pi} \ln\left[ \mathi H (T-T'+X-X'-\mathi \epsilon) \right] \\
&\quad- \frac{1}{4 \pi} \ln\left[ \mathi H (T-T'-X+X'-\mathi \epsilon) \right] + \bigo{H} \eqend{.}
\end{splitequation}
This is identical with the previous result~\eqref{eq:free_scalar_2pf_minkowski}, and shows that the Minkowski limit does not depend on whether we concentrate on the expanding Poincar{\'e} patch or the full de~Sitter spacetime.

\subsection{Vertex operators}

The vertex operators $V_\gamma$ in the free theory are obtained as normal-ordered exponentials of the scalar field. To compute the normal-ordering, we use the BCH formula~\cite{achillesbonfiglioli2012} to write the exponential in the form
\begin{splitequation}
\mathe^{\mathi \gamma \hat{\phi}(\tau,\theta)} &= \mathe^{\mathi \gamma \sum_n \hat{a}_n^\dagger \phi_n^*(\tau,\theta)} \mathe^{\mathi \gamma \sum_n \hat{a}_n \phi_n(\tau,\theta)} \mathe^{- \frac{\gamma^2}{2} \sum_n \abs{\phi_n(\tau,\theta)}^2} \\
&= \mathe^{\mathi \gamma \sum_n \hat{a}_n^\dagger \phi_n^*(\tau,\theta)} \mathe^{\mathi \gamma \sum_n \hat{a}_n \phi_n(\tau,\theta)} \mathe^{- \frac{\gamma^2}{2} \mathi G^+_\alpha(\tau,\theta,\tau,\theta)} \eqend{,}
\end{splitequation}
where all annihilation operators are ordered to the right. For the last term, we compute explicitly that
\begin{splitequation}
\mathe^{- \frac{\gamma^2}{2} \mathi G^+_\alpha(\tau,\theta,\tau,\theta)} &= \exp\left[ - \frac{\gamma^2}{2} \left[ \frac{\alpha^2}{4} + \frac{1}{4 \pi^2 \alpha^2} \left( \tau - \frac{\pi}{2} \right)^2 - \frac{1}{2 \pi} \ln\left( 1 - \mathe^{- \epsilon} \right) \right] \right] \\
&\approx \epsilon^\frac{\gamma^2}{4 \pi} \exp\left[ - \frac{\gamma^2}{2} \left[ \frac{\alpha^2}{4} + \frac{1}{4 \pi^2 \alpha^2} \left( \tau - \frac{\pi}{2} \right)^2 \right] \right] \eqend{,}
\end{splitequation}
where the last line gives the leading result for small $\epsilon$. We thus can define the vertex operators as
\begin{splitequation}
\label{eq:vertex_operator_def}
V_\gamma(\tau,\theta) &\equiv \normord{ \mathe^{\mathi \gamma \hat{\phi}(\tau,\theta)} } = \lim_{\epsilon \to 0^+} \epsilon^{- \frac{\gamma^2}{4 \pi}} \mathe^{\mathi \gamma \hat{\phi}(\tau,\theta)} \\
&= \mathe^{\mathi \gamma \sum_n \hat{a}_n^\dagger \phi_n^*(\tau,\theta)} \mathe^{\mathi \gamma \sum_n \hat{a}_n \phi_n(\tau,\theta)} \exp\left[ - \frac{\gamma^2}{2} \left[ \frac{\alpha^2}{4} + \frac{1}{4 \pi^2 \alpha^2} \left( \tau - \frac{\pi}{2} \right)^2 \right] \right] \eqend{.}
\end{splitequation}
Note that in contrast to Minkowski spacetime, the last factor is explicitly time-dependent. Analogously, for the exponential of the smeared quantum field $\hat{\phi}(f) \equiv \int f(x) \hat{\phi}(x) \sqrt{-g} \total^2 x$ with $f$ a real test function (known as Weyl operators), we obtain the decomposition
\begin{equation}
\label{eq:weyl_normal_ordered}
\mathe^{\mathi \hat{\phi}(f)} = \mathe^{\mathi \sum_n \hat{a}_n^\dagger \phi_n^*(f)} \mathe^{\mathi \sum_n \hat{a}_n \phi_n(f)} \mathe^{- \frac{\mathi}{2} G^+_\alpha(f,f)}
\end{equation}
with the smeared mode functions
\begin{equation}
\phi_n(f) \equiv \int \phi_n(\tau,\theta) \frac{f(\tau,\theta)}{H^2 \sin^2(\tau)} \total \tau \total \theta \eqend{.}
\end{equation}
and the smeared two-point function
\begin{equation}
\label{eq:free_scalar_2pf_smeared}
G^+_\alpha(f,g) \equiv \iint G^+_\alpha(\tau,\theta,\tau',\theta') \frac{f(\tau,\theta) g(\tau',\theta')}{H^4 \sin^2(\tau) \sin^2(\tau')} \total \tau \total \theta \total \tau' \total \theta' \eqend{.}
\end{equation}

Correlation functions involving multiple vertex operators can then easily be computed, using again the BCH formula in the form
\begin{equation}
\mathe^{\mathi \gamma \sum_n \hat{a}_n \phi_n(\tau,\theta)} \mathe^{\mathi \gamma' \sum_n \hat{a}_n^\dagger \phi_n^*(\tau',\theta')} = \mathe^{\mathi \gamma' \sum_n \hat{a}_n^\dagger \phi_n^*(\tau',\theta')} \mathe^{\mathi \gamma \sum_n \hat{a}_n \phi_n(\tau,\theta)} \mathe^{- \gamma \gamma' \mathi G^+_\alpha(\tau,\theta,\tau',\theta')} \eqend{.}
\end{equation}
We obtain the mildly complicated expression
\begin{splitequation}
\label{eq:vertex_correlator}
&\bra{0_\alpha} V_{\gamma_1}(\tau_1,\theta_1) \cdots V_{\gamma_n}(\tau_n,\theta_n) \ket{0_\alpha} \\
&\quad= \mathe^{- \sum_{1 \leq j < k \leq n} \gamma_j \gamma_k \mathi G^+_\alpha(\tau_j,\theta_j,\tau_k,\theta_k)} \prod_{j=1}^n \exp\left[ - \frac{\gamma_j^2}{2} \left[ \frac{\alpha^2}{4} + \frac{1}{4 \pi^2 \alpha^2} \left( \tau_j - \frac{\pi}{2} \right)^2 \right] \right] \\
&\quad= \lim_{\epsilon \to 0^+} \exp\left[ \sum_{1 \leq j < k \leq n} \frac{\gamma_j \gamma_k}{4 \pi} \ln\left[ \left( 1 - \mathe^{- \mathi (\tau_j-\tau_k+\theta_j-\theta_k - \mathi \epsilon)} \right) \left( 1 - \mathe^{- \mathi (\tau_j-\tau_k-\theta_j+\theta_k - \mathi \epsilon)} \right) \right] \right] \\
&\qquad\times \exp\left[ - \frac{\alpha^2}{8} \Bigg( \sum_{j=1}^n \gamma_j \Bigg)^2 - \frac{1}{8 \pi^2 \alpha^2} \Bigg[ \sum_{j=1}^n \gamma_j \Bigg( \tau_j - \frac{\pi}{2} \Bigg) \Bigg]^2 + \frac{\mathi}{4 \pi} \sum_{1 \leq j < k \leq n} \gamma_j \gamma_k \left( \tau_j - \tau_k \right) \right] \eqend{.}
\end{splitequation}
In the limit $\alpha \to \infty$, in which de~Sitter covariance is restored for derivatives of the scalar field, we see that the correlation function~\eqref{eq:vertex_correlator} vanishes unless $\sum_{j=1}^n \gamma_j = 0$. This limit thus also enforces the neutrality condition, which is well-known from Minkowski spacetime, see for example Refs.~\cite{froebcadamuro2022a,swieca1977}. Combining terms and using~\eqref{eq:free_scalar_2pf_log}, for a neutral correlation function it follows that
\begin{splitequation}
\label{eq:vertex_correlator_neutral}
&\bra{0_\alpha} V_{\gamma_1}(\tau_1,\theta_1) \cdots V_{\gamma_n}(\tau_n,\theta_n) \ket{0_\alpha} \Big\rvert_{\raisebox{0.2em}{\scriptsize $\sum_{j=1}^n \gamma_j = 0$}} \\
&\quad= \exp\left[ - \frac{1}{8 \pi^2 \alpha^2} \Bigg( \sum_{j=1}^n \gamma_j \tau_j \Bigg)^2 \right] \lim_{\epsilon \to 0^+} \prod_{1 \leq j < k \leq n} \left[ 2 \cos(\tau_j - \tau_k - \mathi \epsilon) - 2 \cos(\theta_j - \theta_k) \right]^\frac{\gamma_j \gamma_k}{4 \pi} \eqend{,}
\end{splitequation}
and the first term reduces to $1$ in the limit $\alpha \to \infty$.

To obtain correlation functions involving $\hat{\phi}$, we use that $\hat{\phi}(\tau,\theta) = - \mathi \partial_\gamma V_\gamma(\tau,\theta) \big\rvert_{\gamma = 0}$, such that
\begin{splitequation}
&\bra{0_\alpha} V_{\gamma_1}(\tau_1,\theta_1) \cdots V_{\gamma_n}(\tau_n,\theta_n) \hat{\phi}(\tau',\theta') \ket{0_\alpha} \\
&= - \mathi \partial_{\gamma_{n+1}} \bra{0_\alpha} V_{\gamma_1}(\tau_1,\theta_1) \cdots V_{\gamma_n}(\tau_n,\theta_n) V_\gamma(\tau',\theta') \ket{0_\alpha} \big\rvert_{\gamma_{n+1} = 0} \eqend{,}
\end{splitequation}
and then employ the result~\eqref{eq:vertex_correlator} for the correlation function on the right-hand side. Performing the derivative, we can write the result in the form
\begin{splitequation}
\label{eq:vertex_phi_correlator}
&\bra{0_\alpha} V_{\gamma_1}(\tau_1,\theta_1) \cdots V_{\gamma_n}(\tau_n,\theta_n) \hat{\phi}(\tau',\theta') \ket{0_\alpha} \\
&\quad= \bra{0_\alpha} V_{\gamma_1}(\tau_1,\theta_1) \cdots V_{\gamma_n}(\tau_n,\theta_n) \ket{0_\alpha} \Bigg[ \mathi \frac{\alpha^2}{4} \sum_{j=1}^n \gamma_j + \frac{\mathi}{4 \pi^2 \alpha^2} \left( \tau' - \frac{\pi}{2} \right) \sum_{j=1}^n \gamma_j \left( \tau_j - \frac{\pi}{2} \right) \\
&\qquad\quad- \frac{\mathi}{4 \pi} \sum_{j=1}^n \gamma_j \ln\left[ 2 \cos(\tau_j - \tau' - \mathi \epsilon) - 2 \cos(\theta_j - \theta') \right] \Bigg] \eqend{,} \raisetag{2.4em}
\end{splitequation}
where we expressed the remaining exponentials again using the correlator of vertex operators. The result~\eqref{eq:vertex_phi_correlator} simplifies in the limit $\alpha \to \infty$ if the correlator is neutral, $\sum_{j=1}^n \gamma_j = 0$. Otherwise it vanishes if at least one vertex operator is present ($n > 0$). For two scalar fields, we obtain analogously (assuming already the neutrality condition)
\begin{splitequation}
\label{eq:vertex_phiphi_correlator_neutral}
&\bra{0_\alpha} V_{\gamma_1}(\tau_1,\theta_1) \cdots V_{\gamma_n}(\tau_n,\theta_n) \hat{\phi}(\tau',\theta') \hat{\phi}(\tau'',\theta'') \ket{0_\alpha} \Big\rvert_{\raisebox{0.2em}{\scriptsize $\sum_{j=1}^n \gamma_j = 0$}} \\
&\quad\to \bra{0_\alpha} V_{\gamma_1}(\tau_1,\theta_1) \cdots V_{\gamma_n}(\tau_n,\theta_n) \ket{0_\alpha} \Bigg[ \frac{\alpha^2}{4} - \frac{1}{4 \pi} \ln\left[ 2 \cos(\tau' - \tau'' - \mathi \epsilon) - 2 \cos(\theta' - \theta'') \right] \\
&\qquad\quad- \frac{1}{(4 \pi)^2} \sum_{j=1}^n \gamma_j \ln\left[ 2 \cos(\tau_j - \tau' - \mathi \epsilon) - 2 \cos(\theta_j - \theta') \right] \\
&\qquad\qquad\qquad\times \sum_{k=1}^n \gamma_k \ln\left[ 2 \cos(\tau_k - \tau'' - \mathi \epsilon) - 2 \cos(\theta_k - \theta'') \right] \Bigg] \quad (\alpha \to \infty) \eqend{,} \raisetag{2.4em}
\end{splitequation}
which now contains a divergent term. This is of course expected, since in the case of zero vertex operators ($n = 0$) it must reduce to the two-point function~\eqref{eq:free_scalar_2pf} which shows this divergence. Nevertheless, taking a derivative this term disappears, and the limit $\alpha \to \infty$ is well-defined and finite.

However, unlike for the two-point function of derivatives of the scalar field~\eqref{eq:free_scalar_2pf_derivative_boost}, the correlation function of vertex operators~\eqref{eq:vertex_correlator} is not covariant under the boost symmetries, even if the neutrality condition holds and in the limit $\alpha \to \infty$ where the correlator simplifies~\eqref{eq:vertex_correlator_neutral}.

By a similar computation we obtain correlation functions including Weyl operators $\mathe^{\mathi \hat{\phi}(f)}$~\eqref{eq:weyl_normal_ordered}. The ones that will be needed later on are
\begin{splitequation}
\label{eq:vertex_weyl_correlator}
&\bra{0_\alpha} \mathe^{- \mathi \hat{\phi}(f)} V_{\gamma_1}(\tau_1,\theta_1) \cdots V_{\gamma_n}(\tau_n,\theta_n) \, \mathe^{\mathi \hat{\phi}(f)} \ket{0_\alpha} \\
&\quad= \bra{0_\alpha} V_{\gamma_1}(\tau_1,\theta_1) \cdots V_{\gamma_n}(\tau_n,\theta_n) \ket{0_\alpha} \, \mathe^{\mathi \sum_{i=1}^n \gamma_i \left[ G^+_\alpha(f,\tau_i,\theta_i) - G^+_\alpha(\tau_i,\theta_i,f) \right]}
\end{splitequation}
and (assuming already the neutrality condition)
\begin{splitequation}
\label{eq:vertex_phi_phi_weyl_correlator_neutral}
&\bra{0_\alpha} \mathe^{- \mathi \hat{\phi}(f)} \hat{\phi}(\tau',\theta') V_{\gamma_1}(\tau_1,\theta_1) \cdots V_{\gamma_n}(\tau_n,\theta_n) \hat{\phi}(\tau'',\theta'') \, \mathe^{\mathi \hat{\phi}(f)} \ket{0_\alpha} \Big\rvert_{\raisebox{0.2em}{\scriptsize $\sum_{j=1}^n \gamma_j = 0$}} \\
&\quad= \bra{0_\alpha} \mathe^{- \mathi \hat{\phi}(f)} V_{\gamma_1}(\tau_1,\theta_1) \cdots V_{\gamma_n}(\tau_n,\theta_n) \, \mathe^{\mathi \hat{\phi}(f)} \ket{0_\alpha} \Big\rvert_{\raisebox{0.2em}{\scriptsize $\sum_{j=1}^n \gamma_j = 0$}} \Bigg[ \frac{\alpha^2}{4} \\
&\qquad- \frac{1}{4 \pi} \ln\left[ 2 \cos(\tau'-\tau'' - \mathi \epsilon) - 2 \cos(\theta'-\theta'') \right] + \frac{1}{4 \pi^2 \alpha^2} \Bigg( \tau' - \frac{\pi}{2} \Bigg) \Bigg( \tau'' - \frac{\pi}{2} \Bigg) \\
&\qquad+ \bigg[ G^+_\alpha(f,\tau',\theta') - G^+_\alpha(\tau',\theta',f) + \frac{\mathi}{4 \pi^2 \alpha^2} \Bigg( \tau' - \frac{\pi}{2} \Bigg) \sum_{k=1}^n \gamma_k \tau_k \\
&\qquad\qquad- \frac{\mathi}{4 \pi} \sum_{k=1}^n \gamma_k \ln\left[ 2 \cos(\tau'-\tau_k - \mathi \epsilon) - 2 \cos(\theta'-\theta_k) \right] \bigg] \\
&\qquad\quad\times \bigg[ G^+_\alpha(f,\tau'',\theta'') - G^+_\alpha(\tau'',\theta'',f) + \frac{\mathi}{4 \pi^2 \alpha^2} \Bigg( \tau'' - \frac{\pi}{2} \Bigg) \sum_{k=1}^n \gamma_k \tau_k \\
&\qquad\qquad- \frac{\mathi}{4 \pi} \sum_{k=1}^n \gamma_k \ln\left[ 2 \cos(\tau_k-\tau'' - \mathi \epsilon) - 2 \cos(\theta_k-\theta'') \right] \bigg] \Bigg] \eqend{.}
\end{splitequation}

Analogously to the case of the two-point function~\eqref{eq:free_scalar_feynman}, time-ordered correlation functions of vertex operators are obtained from these results by replacing $\tau_j - \tau_k \to \abs{\tau_j - \tau_k}$, at least if $- \gamma_j \gamma_k < 4 \pi$. This condition arises in the following way: we see from~\eqref{eq:vertex_correlator_neutral} that the time-ordered correlation function is always a well-defined distribution outside of any partial diagonal, that is if $(\tau_j,\theta_j) \neq (\tau_k,\theta_k)$ for all $j$ and $k$. If $- \gamma_j \gamma_k < 4 \pi$, the singularity at the coincidence point $(\tau_j,\theta_j) = (\tau_k,\theta_k)$ is integrable, and we can thus extend the time-ordered correlation function by continuity to this partial diagonal. More precisely, using that $2 \cos(x) = 2 - x^2 + \bigo{x^4}$ we read off from the explicit expression~\eqref{eq:vertex_correlator_neutral} that the scaling degree of the distribution at $x_j - x_k = 0$ in the limit $\epsilon \to 0$ is given by $- \frac{\gamma_j \gamma_k}{2 \pi}$. We recall that the scaling degree of a distribution $u \in \mathcal{S}'(\mathbb{R}^d)$ at $x = 0$ is defined as~\cite{brunettifredenhagen2000}
\begin{equation}
\label{eq:scaling_degree}
\operatorname{sd}(u) \equiv \inf\left\{ a \in \mathbb{R}\colon \lim_{\lambda \to 0} \lambda^a u(f_\lambda) = 0 \ \forall f \in \mathcal{S}(\mathbb{R}^d) \right\} \eqend{,}
\end{equation}
where $f_\lambda(x) \equiv f(\lambda x)$. If this is less than the spacetime dimension $d$, that is less than the dimension of the submanifold in which these points lie, then a unique extension to the diagonal $x_j = x_k$ exists~\cite[Thm.~5.2]{brunettifredenhagen2000}. Similarly, the scaling degree at $x_{k_1} = \ldots = x_{k_\ell}$ in the limit $\epsilon \to 0$ is given by
\begin{equation}
- \sum_{1 \leq i < j \leq \ell} \frac{\gamma_{k_i} \gamma_{k_j}}{2 \pi} = \frac{1}{4 \pi} \sum_{i=1}^\ell \gamma_{k_i}^2 - \frac{1}{4 \pi} \left( \sum_{i=1}^\ell \gamma_{k_i} \right)^2 < \ell \eqend{,}
\end{equation}
where the inequality holds if $\gamma_i^2 < 4 \pi$ for all $i$. Since for $\ell \geq 3$ we have $\ell < 2 (\ell-1)$, this is again less than the dimension of the submanifold in which these points lie, and a unique extension to the partial diagonal $x_{k_1} = \ldots = x_{k_\ell}$ exists~\cite[Thm.~5.2]{brunettifredenhagen2000}.

On the other hand, if $- \gamma_j \gamma_k \geq 4 \pi$ such that the scaling degree of the distribution at (at least) one partial diagonal is larger than the dimension of the corresponding submanifold, one needs to perform a proper extension~\cite{duetschfredenhagen1999,brunettifredenhagen2000}, which in physics language corresponds to adding local counterterms depending on the UV cutoff $\epsilon$ and supported at the coincidence point $(\tau_j,\theta_j) = (\tau_k,\theta_k)$. In the following, we will only consider the case $- \gamma_j \gamma_k < 4 \pi$, and can thus perform time ordering by the simple replacement $\tau_j - \tau_k \to \abs{\tau_j - \tau_k}$, and anti-time ordering by the replacement $\tau_j - \tau_k \to - \abs{\tau_j - \tau_k}$. We note that for correlation functions which involve derivatives of the field or composite operators such as the stress tensor, the scaling degree at the partial diagonals increases, and it might be necessary to renormalise even in the case that $- \gamma_j \gamma_k < 4 \pi$ for all $j$ and $k$. For details, we refer to our previous work~\cite{froebcadamuro2022a}, but we do not consider such correlation functions here.

\subsection{Implementing symmetries}
\label{sec:ds_symmetries}

The de~Sitter symmetries (rotations and boosts), which are generated by the Killing vectors~\eqref{eq:desitter_killing_rot} and~\eqref{eq:desitter_killing_boost}, are implemented in the quantum theory by unitary conjugation. The generators of this three-parameter family of unitaries can be obtained via the Noether method, contracting the stress tensor with the corresponding Killing vector and the normal vector to a Cauchy surface $\tau = \text{const}$, and integrating over this surface.

The classical stress tensor for the free massless, minimally coupled scalar field $\phi$ with action~\eqref{eq:free_scalar_action} is given by
\begin{equation}
\label{ew:free_scalar_stresstensor}
T_{\mu\nu}[\phi] = \partial_\mu \phi \partial_\nu \phi - \frac{1}{2} g_{\mu\nu} g^{\rho\sigma} \partial_\rho \phi \partial_\sigma \phi \eqend{,}
\end{equation}
and contracting with the rotation and boost Killing vectors~\eqref{eq:desitter_killing_rot} and~\eqref{eq:desitter_killing_boost} as well as the normal vector $n^\mu \partial_\mu = - \partial_\tau$ and restricting to $\tau = 0$, we obtain
\begin{equations}[eq:free_scalar_stresstensor_contracted]
T_{\mu\nu}[\phi] n^\mu \xi_\text{rot}^\nu \big\rvert_{\tau = 0} &= - \partial_\tau \phi \partial_\theta \phi \big\rvert_{\tau = 0} \eqend{,} \\
T_{\mu\nu}[\phi] n^\mu \xi_\text{boost,1}^\nu \big\rvert_{\tau = 0} &= - \sin(\theta) \partial_\tau \phi \partial_\theta \phi \big\rvert_{\tau = 0} \eqend{,} \\
T_{\mu\nu}[\phi] n^\mu \xi_\text{boost,2}^\nu \big\rvert_{\tau = 0} &= \cos(\theta) \partial_\tau \phi \partial_\theta \phi \big\rvert_{\tau = 0} \eqend{.}
\end{equations}
Replacing now the classical scalar field $\phi$ by the quantized one $\hat\phi$, inserting its mode expansion~\eqref{eq:free_scalar_fieldexpansion} with the modes~\eqref{eq:free_scalar_fockmode}, normal ordering and integrating over the Cauchy surface $\tau = 0$, we obtain the Noether charges
\begin{equations}[eq:noether_charges]
\hat{Q}_\text{rot} &\equiv \int \normord{ T_{\mu\nu}[\hat\phi] } n^\mu \xi_\text{rot}^\nu \big\rvert_{\tau = 0} \total \theta = \sum_{n=1}^\infty n \left( \hat{a}_n^\dagger \hat{a}_n - \hat{a}_{-n}^\dagger \hat{a}_{-n} \right) \eqend{,} \\
\begin{split}
\hat{Q}_\text{boost,1} &\equiv \int \normord{ T_{\mu\nu}[\hat\phi] } n^\mu \xi_\text{boost,1}^\nu \big\rvert_{\tau = 0} \total \theta \\
&= - \frac{\mathi}{2 \alpha \sqrt{4 \pi}} \left( \hat{a}_1 + \hat{a}_1^\dagger + \hat{a}_{-1} + \hat{a}_{-1}^\dagger \right) \hat{a}_0 + \frac{\mathi}{2 \alpha \sqrt{4 \pi}} \hat{a}_0^\dagger \left( \hat{a}_1 + \hat{a}_1^\dagger + \hat{a}_{-1} + \hat{a}_{-1}^\dagger \right) \\
&\quad+ \frac{\mathi}{2} \sum_{n=1}^\infty \sqrt{ n (n+1) } \left( \hat{a}_n^\dagger \hat{a}_{n+1} - \hat{a}_{n+1}^\dagger \hat{a}_n - \hat{a}_{-n-1}^\dagger \hat{a}_{-n} + \hat{a}_{-n}^\dagger \hat{a}_{-n-1} \right) \eqend{,}
\end{split} \\
\begin{split}
\hat{Q}_\text{boost,2} &\equiv \int \normord{ T_{\mu\nu}[\hat\phi] } n^\mu \xi_\text{boost,2}^\nu \big\rvert_{\tau = 0} \total \theta \\
&= \frac{1}{2 \alpha \sqrt{4 \pi}} \left( \hat{a}_1 - \hat{a}_1^\dagger - \hat{a}_{-1} + \hat{a}_{-1}^\dagger \right) \hat{a}_0 - \frac{1}{2 \alpha \sqrt{4 \pi}} \hat{a}_0^\dagger \left( \hat{a}_1 - \hat{a}_1^\dagger - \hat{a}_{-1} + \hat{a}_{-1}^\dagger \right) \\
&\quad- \frac{1}{2} \sum_{n=1}^\infty \sqrt{ n (n+1) } \left( \hat{a}_n^\dagger \hat{a}_{n+1} + \hat{a}_{n+1}^\dagger \hat{a}_n - \hat{a}_{-n}^\dagger \hat{a}_{-n-1} - \hat{a}_{-n-1}^\dagger \hat{a}_{-n} \right) \eqend{.}
\end{split}
\end{equations}
Since the Cauchy surfaces of the global de~Sitter spacetime are compact, it is not necessary to introduce a spatial cutoff which is required in Minkowski spacetime~\cite{requardt1976,duetschfredenhagen1999}.

With the explicit expressions~\eqref{eq:noether_charges}, it is now also clear that the one-parameter family of states $\ket{0_\alpha}$ is not invariant under boosts for any finite $\alpha$. We obtain explicitly
\begin{equations}
\hat{Q}_\text{boost,1} \ket{0_\alpha} &= \frac{\mathi}{2 \alpha \sqrt{4 \pi}} \hat{a}_0^\dagger \left( \hat{a}_1^\dagger + \hat{a}_{-1}^\dagger \right) \ket{0_\alpha} \eqend{,} \\
\hat{Q}_\text{boost,2} \ket{0_\alpha} &= \frac{1}{2 \alpha \sqrt{4 \pi}} \hat{a}_0^\dagger \left( \hat{a}_1^\dagger - \hat{a}_{-1}^\dagger \right) \ket{0_\alpha} \eqend{,}
\end{equations}
which only vanishes in the limit $\alpha \to \infty$. However, this limit is only well-defined for derivatives of the field $\hat{\phi}$, not for the field itself, which is in accordance with the result for the transformation of the two-point function $G^+_\alpha$~\eqref{eq:free_scalar_2pf_boost} and its derivative~\eqref{eq:free_scalar_2pf_derivative_boost} under boosts.

The Noether charges~\eqref{eq:noether_charges} are symmetric operators, and are well-defined on all states of finite particle number. Using the expansion~\eqref{eq:free_scalar_fieldexpansion} of the scalar field, the commutation relations~\eqref{eq:canonical_commutator} and the explicit form of the modes~\eqref{eq:free_scalar_fockmode} and~\eqref{eq:free_scalar_zeromode}, we verify that
\begin{splitequation}
\label{eq:noether_charge_rot_commutator}
\left[ \hat{Q}_\text{rot}, \hat{\phi}(\tau,\theta) \right] &= \sum_{m=1}^\infty m \left[ \hat{a}_m^\dagger \phi_m^*(\tau,\theta) - \hat{a}_m \phi_m(\tau,\theta) + \hat{a}_{-m} \phi_{-m}(\tau,\theta) - \hat{a}_{-m}^\dagger \phi_{-m}^*(\tau,\theta) \right] \\
&= \mathi \partial_\theta \hat{\phi}(\tau,\theta) = \mathi \xi_\text{rot} \hat{\phi}(\tau,\theta) \raisetag{1.6em}
\end{splitequation}
and for $k = 1,2$ also
\begin{equation}
\label{eq:noether_charge_boost_commutator}
\left[ \hat{Q}_\text{boost,k}, \hat{\phi}(\tau,\theta) \right] = \mathi \xi_\text{boost,k} \hat{\phi}(\tau,\theta) \eqend{.}
\end{equation}
Exponentiating, we obtain the three-parameter family of unitaries
\begin{equation}
\label{eq:ds_symmetries_unitaries}
U_{abc} \equiv \exp\left[ \mathi \left( a \hat{Q}_\text{rot} + b \hat{Q}_\text{boost,1} + c \hat{Q}_\text{boost,2} \right) \right] \eqend{,}
\end{equation}
whose action on the field reads
\begin{equation}
\label{eq:ds_symmetries_adjointaction_bare}
U_{abc} \, \hat{\phi}(\tau,\theta) U_{abc}^\dagger = \mathe^{- a \xi_\text{rot} - b \xi_\text{boost,1} - c \xi_\text{boost,2}} \hat{\phi}(\tau,\theta) = \hat{\phi}(\tau_{-a,-b,-c},\theta_{-a,-b,-c})
\end{equation}
with the transformed coordinates~\eqref{eq:appendix_ds_coordinates}, where we used the result~\eqref{eq:desitter_finite_transformation}. For the smeared field $\hat{\phi}(f)$, using the transformation~\eqref{eq:appendix_ds_measure} of the integration measure and the inverse transformation~\eqref{eq:appendix_ds_finite_transformation_inv}, this reads
\begin{splitequation}
\label{eq:ds_symmetries_adjointaction}
U_{abc} \, \hat{\phi}(f) U_{abc}^\dagger &= \int f(x) \hat{\phi}(x_{-a,-b,-c}) \sqrt{-g} \total^2 x \\
&= \int f(x) \hat{\phi}(x_{-a,-b,-c}) \sqrt{-g(x_{-a,-b,-c})} \total^2 x_{-a,-b,-c} \\
&= \int f(x_{abc}) \hat{\phi}(x) \sqrt{-g(x)} \total^2 x = \hat{\phi}(f_{abc})
\end{splitequation}
with the transformed smearing function $f_{abc}(x) = f(x_{abc})$.

Since the span of states with finite particle number is dense in the Fock space~\cite[Thm.~X.41]{reedsimon2}, the unitaries~\eqref{eq:ds_symmetries_unitaries} can then be extended to the full Fock space, and their generators define a self-adjoint operator. Moreover, their action is also well-defined on the vertex operators $V_\gamma$~\eqref{eq:vertex_operator_def}, and analogously to~\eqref{eq:ds_symmetries_adjointaction_bare} we obtain
\begin{equation}
\label{eq:ds_symmetries_adjointaction_vertex}
U_{abc} V_\gamma(\tau,\theta) U_{abc}^\dagger = V_\gamma(\tau_{-a,-b,-c},\theta_{-a,-b,-c}) \eqend{.}
\end{equation}

\section{The Sine--Gordon model and the \texorpdfstring{$S$}{S} matrix}
\label{sec:smatrix}

The classical action for the Sine--Gordon model is given by~\eqref{eq:intro_sinegordon_action}
\begin{equation}
\label{eq:sine_gordon_action}
S = \int \left[ - \frac{1}{2} \nabla^\mu \phi \nabla_\mu \phi + 2 g \cos(\beta \phi) \right] \sqrt{-g} \total^2 x \eqend{,}
\end{equation}
where $\beta > 0$ is the coupling constant and $g$ the interaction cutoff. A priori, one takes $g$ to be a function of compact support, avoiding infrared issues in the construction of the perturbative series. At the end, one is interested in the adiabatic limit $g \to \text{const}$, which we will do in the construction of the net of algebras.

\subsection{Convergent perturbative series for the \texorpdfstring{$S$}{S} matrix}
\label{sec:smatrix_series}

The $S$ matrix is defined by the Bogoliubov formula
\begin{equation}
\label{eq:smatrix_def}
\hat{S}(g) \equiv \mathcal{T}\left[ \mathe_\otimes^{\mathi S_\text{int}} \right] \eqend{,}
\end{equation}
where $S_\text{int}$ is the interaction part of the classical action~\eqref{eq:sine_gordon_action} and $\mathcal{T}$ denotes a time-ordered product, which is a multilinear map from classical fields to the algebra of quantum fields. In general, the definition~\eqref{eq:smatrix_def} needs to be understood as a formal power series in $g$. However, as in Minkowski spacetime~\cite{bahnsrejzner2018,bahnsfredenhagenrejzner2021,bahnspinamontirejzner2023,froebcadamuro2022a} we will show that the series is actually convergent for all $\beta^2 < 4 \pi$.

Writing the interaction as $2 g \cos(\beta \phi) = g \left[ \mathe^{\mathi \beta \phi} + \mathe^{- \mathi \beta \phi} \right]$, the $S$ matrix~\eqref{eq:smatrix_def} is the series $\hat{S}(g) = \sum_{k=0}^\infty \hat{S}_k(g)$ with
\begin{equation}
\label{eq:smatrix_series_def}
\hat{S}_k(g) \equiv \frac{\mathi^k}{k!} \sum_{\sigma_1, \ldots, \sigma_k = \pm 1} \int\dotsi\!\int \mathcal{T}\left[ \bigotimes_{i=1}^k \mathe^{\mathi \sigma_i \beta \phi(\tau_i,\theta_i)} \right] \prod_{i=1}^k \frac{g(\tau_i,\theta_i)}{H^2 \sin^2(\tau_i)} \total \tau_i \total \theta_i \eqend{.}
\end{equation}
The time-ordered products can be constructed inductively, starting with a single entry where the time-ordered product just results in the normal-ordered quantum field:
\begin{equation}
\label{eq:time_ordered_product_single}
\mathcal{T}\left[ \phi \right] = \hat{\phi} \eqend{,} \quad \mathcal{T}\left[ \mathe^{\mathi \beta \phi} \right] = V_\beta \eqend{.}
\end{equation}
Time-ordered products with multiple entries are already well-defined outside of the diagonal by imposing the causal factorization condition:
\begin{equation}
\label{eq:time_ordered_product_causalfactorization}
\mathcal{T}\left[ \op_1(x_1) \otimes \cdots \otimes \op_n(x_n) \right] = \mathcal{T}\left[ \op_1(x_1) \otimes \cdots \otimes \op_k(x_k) \right] \mathcal{T}\left[ \op_{k+1}(x_{k+1}) \otimes \cdots \otimes \op_n(x_n) \right]
\end{equation}
if all $x_i$ with $i \in \{1,\ldots,k\}$ do not lie in the future of any of the $x_i$ with $i \in \{k+1,\ldots,n\}$, i.e., they are in the past of or spacelike separated from them. The extension to the diagonal then corresponds to renormalization and defines the time-ordered products on all of spacetime. In our case, taking $\beta^2 < 4 \pi$, the extension can be done by continuity and no renormalization is needed. In the same way, one recursively constructs anti-time-ordered products $\overline{\mathcal{T}}$, which instead of the causal factorization condition~\eqref{eq:time_ordered_product_causalfactorization} fulfill anti-causal factorization
\begin{equation}
\label{eq:anti_time_ordered_product_causalfactorization}
\overline{\mathcal{T}}\left[ \op_1(x_1) \otimes \cdots \otimes \op_n(x_n) \right] = \overline{\mathcal{T}}\left[ \op_1(x_1) \otimes \cdots \otimes \op_k(x_k) \right] \overline{\mathcal{T}}\left[ \op_{k+1}(x_{k+1}) \otimes \cdots \otimes \op_n(x_n) \right]
\end{equation}
if all $x_i$ with $i \in \{1,\ldots,k\}$ do not lie in the past of any of the $x_i$ with $i \in \{k+1,\ldots,n\}$, i.e., they are in the future of or spacelike separated from them.

We will show that the series $\sum_{k=0}^\infty \hat{S}_k(g)$ converges strongly on a dense domain $D$ in Fock space, namely on the span of the vectors $\mathe^{\mathi \hat{\phi}(f)} \ket{0_\alpha}$~\cite[Prop.~5.2.4]{brattelirobinson2}, where $\hat{\phi}(f) \equiv \int f(x) \hat{\phi}(x) \sqrt{-g} \total^2 x$ is the quantum field smeared with a real test function $f$. We will moreover show that the sum is unitary and defines a bounded operator on this dense domain. It follows that $\hat{S}(g)$ has a unique extension to the full Fock space, and defines a bounded operator there.

We thus have to bound the norm
\begin{splitequation}
\label{eq:smatrix_norm_expansion}
\norm{ \hat{S}_k(g) \mathe^{\mathi \hat{\phi}(f)} \ket{0_\alpha} } &\leq \frac{1}{k!} \sum_{\sigma_1, \ldots, \sigma_k = \pm 1} \\
&\qquad \norm{ \int\dotsi\!\int \mathcal{T}\left[ \bigotimes_{i=1}^k \mathe^{\mathi \sigma_i \beta \phi(\tau_i,\theta_i)} \right] \prod_{i=1}^k \frac{g(\tau_i,\theta_i)}{H^2 \sin^2(\tau_i)} \total \tau_i \total \theta_i \, \mathe^{\mathi \hat{\phi}(f)} \ket[\Big]{0_\alpha} } \eqend{.}
\end{splitequation}
Using the correlation functions~\eqref{eq:vertex_weyl_correlator}, we obtain an expression for
\begin{splitequation}
\label{eq:smatrix_term_normsquared}
&\norm{ \int\dotsi\!\int \mathcal{T}\left[ \bigotimes_{i=1}^k \mathe^{\mathi \sigma_i \beta \phi(\tau_i,\theta_i)} \right] \prod_{i=1}^k \frac{g(\tau_i,\theta_i)}{H^2 \sin^2(\tau_i)} \total \tau_i \total \theta_i \, \mathe^{\mathi \hat{\phi}(f)} \ket[\Big]{0_\alpha} }^2 \\
&\quad= \int\dotsi\!\int \bra[\Big]{0_\alpha} \mathe^{- \mathi \hat{\phi}(f)} \overline{\mathcal{T}}\left[ \bigotimes_{i=1}^k \mathe^{- \mathi \sigma_i \beta \phi(\tau_i,\theta_i)} \right] \mathcal{T}\Bigg[ \bigotimes_{i=k+1}^{2k} \mathe^{\mathi \sigma_{i-k} \beta \phi(\tau_i,\theta_i)} \Bigg] \mathe^{\mathi \hat{\phi}(f)} \ket[\Big]{0_\alpha} \\
&\qquad\quad\times \prod_{i=1}^{2k} \frac{g(\tau_i,\theta_i)}{H^2 \sin^2(\tau_i)} \total \tau_i \total \theta_i \eqend{,}
\end{splitequation}
where we used that the adjoint of a time-ordered product is an anti-time-ordered product, and where we replace all differences $\tau_j - \tau_k$ by $\abs{\tau_j - \tau_k}$ for operators appearing in the time-ordered product $\mathcal{T}$ and by $- \abs{\tau_j - \tau_k}$ for operators appearing in the anti-time-ordered product $\overline{\mathcal{T}}$ to implement the (anti-)time-ordering. Since the correlation function~\eqref{eq:smatrix_term_normsquared} fulfills the neutrality condition (because every vertex operator appears twice, once with $\sigma_i \beta$ and once with $- \sigma_i \beta$), we can then use the result~\eqref{eq:vertex_correlator_neutral} and obtain
\begin{splitequation}
\label{eq:smatrix_term_normsquared_2}
&\bra[\Big]{0_\alpha} \mathe^{- \mathi \hat{\phi}(f)} \overline{\mathcal{T}}\left[ \bigotimes_{i=1}^k \mathe^{- \mathi \sigma_i \beta \phi(\tau_i,\theta_i)} \right] \mathcal{T}\Bigg[ \bigotimes_{i=k+1}^{2k} \mathe^{\mathi \sigma_i \beta \phi(\tau_i,\theta_i)} \Bigg] \mathe^{\mathi \hat{\phi}(f)} \ket[\Big]{0_\alpha} \\
&\quad= \lim_{\epsilon \to 0^+} \prod_{1 \leq i < j \leq k} \left[ 2 \cos( - \abs{\tau_i - \tau_j} - \mathi \epsilon) - 2 \cos(\theta_i - \theta_j) \right]^\frac{\sigma_i \sigma_j \beta^2}{4 \pi} \\
&\qquad\quad\times \prod_{i=1}^k \prod_{j=k+1}^{2k} \left[ 2 \cos(\tau_i - \tau_j - \mathi \epsilon) - 2 \cos(\theta_i - \theta_j) \right]^\frac{\sigma_i \sigma_{j-k} \beta^2}{4 \pi} \\
&\qquad\quad\times \prod_{k+1 \leq i < j \leq 2k} \left[ 2 \cos( \abs{\tau_i - \tau_j} - \mathi \epsilon) - 2 \cos(\theta_i - \theta_j) \right]^\frac{\sigma_{i-k} \sigma_{j-k} \beta^2}{4 \pi} \\
&\qquad\quad\times \mathe^{- \frac{\beta^2}{8 \pi^2 \alpha^2} \left[ \sum_{j=1}^k \sigma_j ( \tau_j - \tau_{j+k} ) \right]^2} \\
&\qquad\quad\times \mathe^{\mathi \sum_{i=1}^k \sigma_i \beta \left[ - G^+_\alpha(f,\tau_i,\theta_i) + G^+_\alpha(\tau_i,\theta_i,f) + G^+_\alpha(f,\tau_{i+k},\theta_{i+k}) - G^+_\alpha(\tau_{i+k},\theta_{i+k},f) \right]}
\end{splitequation}
To bound this expression, we first consider the last terms. Using the explicit expression~\eqref{eq:free_scalar_2pf} of the two-point function together with~\eqref{eq:free_scalar_2pf_log}, we compute
\begin{splitequation}
&G^+_\alpha(f,\tau,\theta) - G^+_\alpha(\tau,\theta,f) = \int \left[ G^+_\alpha(\tau',\theta',\tau,\theta) - G^+_\alpha(\tau,\theta,\tau',\theta') \right] \frac{f(\tau',\theta')}{H^2 \sin^2(\tau')} \total \tau' \total \theta' \\
&\qquad= - \int \sgn[ \sin(\tau-\tau') ] \Theta\left[ \cos(\theta-\theta') - \cos(\tau-\tau') \right] \frac{f(\tau',\theta')}{2 H^2 \sin^2(\tau')} \total \tau' \total \theta' \eqend{,}
\end{splitequation}
and since this is real, the last term in~\eqref{eq:smatrix_term_normsquared_2} is just a phase factor and bounded by $1$. The second-to-last term is clearly also bounded by $1$, and in fact equal to $1$ in the limit $\alpha \to \infty$.

For the remaining terms in~\eqref{eq:smatrix_term_normsquared_2}, we first note that for $\beta < 4 \pi$ the singularities at the partial diagonals are integrable, and we can thus take the limit $\epsilon \to 0^+$ in each term. They then have all the same structure, and we can simplify the resulting expression by renaming integration variables. Namely, replacing $(\tau_i,\theta_i) \to (\tilde{\tau}_i,\tilde{\theta}_i)$ for all $i$ with $\sigma_i = -1$, we obtain
\begin{splitequation}
\label{eq:smatrix_term_normsquared_3}
&\norm{ \int\dotsi\!\int \mathcal{T}\left[ \bigotimes_{i=1}^k \mathe^{\mathi \sigma_i \beta \phi(\tau_i,\theta_i)} \right] \prod_{i=1}^k \frac{g(\tau_i,\theta_i)}{H^2 \sin^2(\tau_i)} \total \tau_i \total \theta_i \, \mathe^{\mathi \hat{\phi}(f)} \ket[\Big]{0_\alpha} }^2 \\
&\quad\leq \int\dotsi\!\int \abs{ \frac{\prod_{1 \leq i < j \leq k} \Big[ 2 \cos(\tau_i - \tau_j) - 2 \cos(\theta_i - \theta_j) \Big] \left[ 2 \cos(\tilde{\tau}_i - \tilde{\tau}_j) - 2 \cos(\tilde{\theta}_i - \tilde{\theta}_j) \right]}{\prod_{i,j=1}^k \left[ 2 \cos(\tau_i - \tilde{\tau}_j) - 2 \cos(\theta_i - \tilde{\theta}_j) \right]} }^\frac{\beta^2}{4 \pi} \\
&\qquad\qquad\times \prod_{i=1}^k \frac{\abs{ g(\tau_i,\theta_i) g(\tilde{\tau}_i,\tilde{\theta}_i) }}{H^4 \sin^2(\tau_i) \sin^2(\tilde{\tau}_i)} \total \tau_i \total \theta_i \total \tilde{\tau}_i \total \tilde{\theta}_i \eqend{.} \raisetag{2em}
\end{splitequation}
This expression is very similar to the flat-space one (see, e.g.,~\cite[Eq.~(159)]{froebcadamuro2022a}), with differences between time and space coordinates replaced by the differences between the cosines. We therefore can also estimate it using similar methods, and compute first that
\begin{equations}[eq:cosine_identities]
\abs{ 2 \cos(\tau_i - \tau_j) - 2 \cos(\theta_i - \theta_j) }^2 &= \left[ 2 - 2 \cos(u_i - u_j) \right] \left[ 2 - 2 \cos(v_i - v_j) \right] \eqend{,} \label{eq:cosine_identity_1} \\
2 - 2 \cos(x-y) &= \abs{ \mathe^{\mathi x} - \mathe^{\mathi y} }^2 \label{eq:cosine_identity_2}
\end{equations}
with
\begin{equation}
\label{eq:light_cone_coordinates_def}
u_i \equiv \tau_i - \theta_i \in (-2\pi, \pi) \eqend{,} \quad v_i \equiv \tau_i + \theta_i \in [0, 3\pi) \eqend{.}
\end{equation}
Changing integration variables to $u_i$ and $v_i$ using that $\total \tau_i \total \theta_i = \frac{1}{2} \total u_i \total v_i$, it follows that
\begin{splitequation}
\label{eq:smatrix_term_normsquared_4}
&\norm{ \int\dotsi\!\int \mathcal{T}\left[ \bigotimes_{i=1}^k \mathe^{\mathi \sigma_i \beta \phi(\tau_i,\theta_i)} \right] \prod_{i=1}^k \frac{g(\tau_i,\theta_i)}{H^2 \sin^2(\tau_i)} \total \tau_i \total \theta_i \, \mathe^{\mathi \hat{\phi}(f)} \ket[\Big]{0_\alpha} }^2 \\
&\quad\leq \int\dotsi\!\int \left[ \frac{\prod_{1 \leq i < j \leq k} \abs{ \mathe^{\mathi u_i} - \mathe^{\mathi u_j} } \abs{ \mathe^{\mathi \tilde{u}_i} - \mathe^{\mathi \tilde{u}_j} }}{\prod_{i,j=1}^k \abs{ \mathe^{\mathi u_i} - \mathe^{\mathi \tilde{u}_j} }} \right]^\frac{\beta^2}{4 \pi} \left[ \frac{\prod_{1 \leq i < j \leq k} \abs{ \mathe^{\mathi v_i} - \mathe^{\mathi v_j} } \abs{ \mathe^{\mathi \tilde{v}_i} - \mathe^{\mathi \tilde{v}_j} }}{\prod_{i,j=1}^k \abs{ \mathe^{\mathi v_i} - \mathe^{\mathi \tilde{v}_j} }} \right]^\frac{\beta^2}{4 \pi} \\
&\qquad\qquad\times \prod_{i=1}^k \frac{\abs{ g(u_i,v_i) g(\tilde{u}_i,\tilde{v}_i) }}{4 H^4 \sin^2\left( \frac{u_i+v_i}{2} \right) \sin^2\left( \frac{\tilde{u}_i + \tilde{v}_i}{2} \right)} \prod_{i=1}^k \total u_i \total \tilde{u}_i \total v_i \total \tilde{v}_i \eqend{.} \raisetag{2em}
\end{splitequation}
We see that the integrand factorizes into terms depending on $u$, terms depending on $v$, and terms depending on the adiabatic cutoff $g$. We now use the Hölder inequality
\begin{equation}
\label{eq:hoelder_inequality}
\norm{ f g }_1 \leq \norm{ f }_\frac{1}{p} \norm{ g }_\frac{1}{1-p} \eqend{,} \quad p \in (0,1)
\end{equation}
and choose $p = (4\pi+\beta^2)/(8\pi) < 1$ to obtain
\begin{splitequation}
\label{eq:smatrix_term_normsquared_5}
&\norm{ \int\dotsi\!\int \mathcal{T}\left[ \bigotimes_{i=1}^k \mathe^{\mathi \sigma_i \beta \phi(\tau_i,\theta_i)} \right] \prod_{i=1}^k \frac{g(\tau_i,\theta_i)}{H^2 \sin^2(\tau_i)} \total \tau_i \total \theta_i \, \mathe^{\mathi \hat{\phi}(f)} \ket[\Big]{0_\alpha} }^2 \\
&\quad\leq \left[ \int\dotsi\!\int \left[ \frac{\prod_{1 \leq i < j \leq k} \abs{ \mathe^{\mathi u_i} - \mathe^{\mathi u_j} } \abs{ \mathe^{\mathi \tilde{u}_i} - \mathe^{\mathi \tilde{u}_j} }}{\prod_{i,j=1}^k \abs{ \mathe^{\mathi u_i} - \mathe^{\mathi \tilde{u}_j} }} \right]^\frac{2 \beta^2}{4 \pi + \beta^2} \prod_{i=1}^k \total u_i \total \tilde{u}_i \right]^\frac{4 \pi + \beta^2}{8 \pi} \\
&\qquad\times \left[ \int\dotsi\!\int \left[ \frac{\prod_{1 \leq i < j \leq k} \abs{ \mathe^{\mathi v_i} - \mathe^{\mathi v_j} } \abs{ \mathe^{\mathi \tilde{v}_i} - \mathe^{\mathi \tilde{v}_j} }}{\prod_{i,j=1}^k \abs{ \mathe^{\mathi v_i} - \mathe^{\mathi \tilde{v}_j} }} \right]^\frac{2 \beta^2}{4 \pi + \beta^2} \prod_{i=1}^k \total v_i \total \tilde{v}_i \right]^\frac{4 \pi + \beta^2}{8 \pi} \\
&\qquad\times \norm{ \frac{g(\tau,\theta)}{2 H^2 \sin^2(\tau)} }_{\frac{8 \pi}{4 \pi - \beta^2}}^{2k} \eqend{,}
\end{splitequation}
where the norm is defined by (changing integration variables back to $\tau$ and $\theta$)
\begin{equation}
\norm{ \frac{g(\tau,\theta)}{2 H^2 \sin^2(\tau)} }_p \equiv \left[ 2 \iint \abs{ \frac{g(\tau,\theta)}{2 H^2 \sin^2(\tau)} }^p \total \tau \total \theta \right]^\frac{1}{p} \eqend{.}
\end{equation}
We note that the singularity in the remaining integrals in~\eqref{eq:smatrix_term_normsquared_5} is still integrable since $\frac{2 \beta^2}{4 \pi + \beta^2} < 1$ for $\beta^2 < 4 \pi$. Moreover, the integrands in $u$ and $v$ are identical, and we thus only need to bound one of them explicitly. For this, we use the Cauchy determinant formula~\cite{cauchy1841}
\begin{equation}
\det\left( \frac{1}{x_i - y_j} \right)_{i,j=1}^k = \frac{\prod_{1 \leq i < j \leq k} (x_i-x_j) (y_i-y_j)}{\prod_{i,j=1}^k (x_i-y_j)}
\end{equation}
to bound
\begin{splitequation}
\label{eq:cauchy_bound}
\left[ \frac{\prod_{1 \leq i < j \leq k} \abs{ \mathe^{\mathi v_i} - \mathe^{\mathi v_j} } \abs{ \mathe^{\mathi \tilde{v}_i} - \mathe^{\mathi \tilde{v}_j} }}{\prod_{i,j=1}^k \abs{ \mathe^{\mathi v_i} - \mathe^{\mathi \tilde{v}_j} }} \right]^p &= \abs{ \det\left( \frac{1}{\mathe^{\mathi v_i} - \mathe^{\mathi \tilde{v}_j}} \right)_{i,j=1}^k }^p \\
&\leq \abs{ \sum_\pi \prod_{i=1}^k \frac{1}{\abs{\mathe^{\mathi v_i} - \mathe^{\mathi \tilde{v}_{\pi(i)}}}} }^p \\
&\leq \sum_\pi \prod_{i=1}^k \abs{\mathe^{\mathi v_i} - \mathe^{\mathi \tilde{v}_{\pi(i)}}}^{-p} \eqend{,}
\end{splitequation}
where the sum runs over all permutations $\pi$ of $\{1,\ldots,k\}$, and we have used the inequality
\begin{equation}
\left( \sum_{j=1}^k \abs{a_k} \right)^p \leq \sum_{j=1}^k \abs{a_k}^p \eqend{,} \quad 0 < p < 1
\end{equation}
with $p = 2 \beta^2/(4 \pi + \beta^2) < 1$ to take the sum out of the absolute value.

Inserting the bound~\eqref{eq:cauchy_bound} into the estimate~\eqref{eq:smatrix_term_normsquared_5}, replacing the sum over permutations by a factor $k!$ (since all permutations contribute the same amount), and simplifying terms, it follows that
\begin{splitequation}
\label{eq:smatrix_term_normsquared_6}
&\norm{ \int\dotsi\!\int \mathcal{T}\left[ \bigotimes_{i=1}^k \mathe^{\mathi \sigma_i \beta \phi(\tau_i,\theta_i)} \right] \prod_{i=1}^k \frac{g(\tau_i,\theta_i)}{H^2 \sin^2(\tau_i)} \total \tau_i \total \theta_i \, \mathe^{\mathi \hat{\phi}(f)} \ket[\Big]{0_\alpha} }^2 \\
&\quad\leq \norm{ \frac{g(\tau,\theta)}{2 H^2 \sin^2(\tau)} }_{\frac{8 \pi}{4 \pi - \beta^2}}^{2k} (k!)^\frac{4 \pi + \beta^2}{4 \pi} \\
&\qquad\times \left[ \iint \abs{\mathe^{\mathi u} - \mathe^{\mathi \tilde{u}}}^{-\frac{2 \beta^2}{4 \pi + \beta^2}} \total u \total \tilde{u} \right]^{k \frac{4 \pi + \beta^2}{8 \pi}} \left[ \iint \abs{\mathe^{\mathi v} - \mathe^{\mathi \tilde{v}}}^{-\frac{2 \beta^2}{4 \pi + \beta^2}} \total v \total \tilde{v} \right]^{k \frac{4 \pi + \beta^2}{8 \pi}} \eqend{.}
\end{splitequation}
To finally bound the integrals, we use again the identity~\eqref{eq:cosine_identity_2}, and then the bound~\eqref{eq:appendix_cosine_bound}
\begin{equation}
\frac{1}{2 - 2 \cos(x)} \leq \begin{cases} \frac{\pi^2}{8 x^2} & x \in \left[ - \frac{\pi}{2}, \frac{\pi}{2} \right] \\ \frac{1}{2} & x \in \left[ \frac{\pi}{2}, \frac{3 \pi}{2} \right] \eqend{.} \end{cases}
\end{equation}
Taking into account that $u, \tilde{u} \in (-2\pi,\pi)$, it follows that
\begin{splitequation}
&\iint \abs{\mathe^{\mathi u} - \mathe^{\mathi \tilde{u}}}^{-\frac{2 \beta^2}{4 \pi + \beta^2}} \total u \total \tilde{u} \leq 3 \pi \int_{-4\pi}^{2\pi} \left[ 2 - 2 \cos(u) \right]^{-\frac{\beta^2}{4 \pi + \beta^2}} \total u \\
&\qquad\leq 9 \pi \left[ \int_0^\frac{\pi}{2} \left( \frac{\pi^2}{8 u^2} \right)^\frac{\beta^2}{4 \pi + \beta^2} \total u + \int_\frac{\pi}{2}^\frac{3\pi}{2} 2^{-\frac{\beta^2}{4 \pi + \beta^2}} \total u + \int_\frac{3\pi}{2}^{2\pi} \left( \frac{\pi^2}{8 (u-2\pi)^2} \right)^\frac{\beta^2}{4 \pi + \beta^2} \total u \right] \\
&\qquad= \frac{36 \cdot 2^\frac{4\pi}{4\pi+\beta^2} \pi^3}{4 \pi - \beta^2} \leq \frac{72 \pi^3}{4 \pi - \beta^2} \eqend{,}
\end{splitequation}
where we used the $2\pi$-periodicity of the cosine to reduce the integration to the interval $[0,2\pi]$. The same bound is obtained for the integral over $v$ and $\tilde{v}$. Taking all together and using some straightforward further estimates, we obtain
\begin{equation}
\label{eq:smatrix_term_normsquared_erg}
\norm{ \int\dotsi\!\int \mathcal{T}\left[ \bigotimes_{i=1}^k \mathe^{\mathi \sigma_i \beta \phi(\tau_i,\theta_i)} \right] \prod_{i=1}^k \frac{g(\tau_i,\theta_i)}{H^2 \sin^2(\tau_i)} \total \tau_i \total \theta_i \, \mathe^{\mathi \hat{\phi}(f)} \ket[\Big]{0_\alpha} }^2 \leq (k!)^{1 + \frac{\beta^2}{4 \pi}} C(g)^{2k}
\end{equation}
with the constant
\begin{equation}
\label{eq:smatrix_cg_def}
C(g) \equiv \frac{36 \sqrt{2} \, \pi^3}{(4 \pi - \beta^2)^\frac{4 \pi + \beta^2}{8 \pi}} \, \norm{ \frac{g(\tau,\theta)}{2 H^2 \sin^2(\tau)} }_{\frac{8 \pi}{4 \pi - \beta^2}} \eqend{.}
\end{equation}

For the norm~\eqref{eq:smatrix_norm_expansion}, we thus obtain the bound
\begin{equation}
\norm{ \hat{S}_k(g) \mathe^{\mathi \hat{\phi}(f)} \ket{0_\alpha} } \leq (k!)^{- \frac{4 \pi - \beta^2}{8 \pi}} [ 2 C(g) ]^k \eqend{,}
\end{equation}
where the factor $2^k$ comes from the sum over the $\sigma_i = \pm 1$. For $\beta^2 < 4 \pi$, this is summable in $k$ for any adiabatic cutoff $g$ such that $C(g) < \infty$, which holds whenever the support of $g$ does not include $\tau = 0$ nor $\tau = \pi$. It follows that the sum $\sum_{k=0}^\infty \hat{S}_k(g)$ converges strongly on the vectors $\mathe^{\mathi \hat{\phi}(f)} \ket{0_\alpha}$, and by extension on any finite linear combination. That is, it converges strongly on the dense domain $D$ in Fock space, such that the S matrix $\hat{S}(g)$ is a well-defined operator on this domain.

Following~\cite{bahnsfredenhagenrejzner2021}, we also show that the S matrix $\hat{S}(g)$ is a unitary operator. It is formally unitary, i.e., we have
\begin{equation}
\sum_{k=0}^n \hat{S}_k(g) \hat{S}_{n-k}^\dagger(g) = \sum_{k=0}^n \hat{S}^\dagger_k(g) \hat{S}_{n-k}(g) = \delta_{n,0} \eqend{,}
\end{equation}
which follows from the definition~\eqref{eq:smatrix_series_def} and the properties of time-ordered products. Since the series defining $\hat{S}(g)$ converges strongly on $D$, the Cauchy product series of the series of $\hat{S}^\dagger(g)$ and $\hat{S}(g)$ converges to the product $\hat{S}^\dagger(g) \hat{S}(g)$, and we have
\begin{equation}
\norm{ \hat{S}(g) \ket{\Psi} }^2 = \sum_{n=0}^\infty \sum_{k=0}^n \bra{\Psi} \hat{S}^\dagger_k(g) \hat{S}_{n-k}(g) \ket{\Psi} = \norm{ \ket{\Psi} }^2
\end{equation}
for any $\ket{\Psi} \in D$. Therefore, $\hat{S}(g)$ is an isometry on the domain $D$, and by continuity has a unique extension to the full Fock space. Analogous arguments hold for $\hat{S}^\dagger(g)$, such that $\hat{S}(g)$ is a unitary operator on the full Fock space.

Lastly, we consider the transformation of the S matrix under rotations and boosts, which is obtained by conjugation with $U_{abc}$. For the $k$-th order term~\eqref{eq:smatrix_series_def}, we use the fact that for $\beta^2 < 4 \pi$ no renormalization is needed and the time-ordered products thus can be extended by continuity to the total diagonal, such that we can use the transformation~\eqref{eq:ds_symmetries_adjointaction_vertex} of the vertex operators $V_{\pm \beta}$. Transforming the integration measure according to~\eqref{eq:appendix_ds_measure}, it is easy to see that
\begin{equation}
U_{abc} \, \hat{S}_k(g) U_{abc}^\dagger = \hat{S}_k(g_{abc})
\end{equation}
with the transformed adiabatic cutoff $g_{abc}(x) \equiv g(x_{abc})$ and the transformed coordinates $x_{abc}$~\eqref{eq:appendix_ds_finite_transformation}. As long as the parameters $a$, $b$, $c$ are finite, the support of $g_{abc}$ does not include $\tau = 0$ nor $\tau = \pi$ if the support of $g$ does not include them, and thus the constant $C(g_{abc})$~\eqref{eq:smatrix_cg_def} for the transformed adiabatic cutoff is finite if $C(g)$ is finite. It follows that the series $\sum_{k=0}^\infty \hat{S}_k(g_{abc})$ is convergent as well, and thus we obtain
\begin{equation}
\label{eq:smatrix_symmetry_transform}
U_{abc} \, \hat{S}(g) U_{abc}^\dagger = \hat{S}(g_{abc})
\end{equation}
for the full S matrix.

\subsection{The Haag--Kastler net}
\label{sec:smatrix_net}

The interacting local net of the Sine--Gordon model is constructed from the relative S matrices, which are defined as~\cite{brunettifredenhagen2000}
\begin{equation}
\label{eq:relative_smatrix_def}
\hat{S}_g(f) \equiv \hat{S}(g)^{-1} \hat{S}(g+f) \eqend{.}
\end{equation}
Since the S matrices $\hat{S}(g)$ are unitary operators for any (real) test function $g$ with $C(g) < \infty$~\eqref{eq:smatrix_cg_def}, one sees easily that also the relative S matrices are unitary. Furthermore, a change of the adiabatic cutoff function $g$ outside of the causal hull\footnote{The causal hull of a set $M$ is the intersection of its causal future and past, $J^+(M) \cap J^-(M)$.} of $\supp f$ only results in a unitary conjugation of the relative S matrix~\cite{duetschfredenhagen1999,brunettifredenhagen2000}. We can therefore perform the adiabatic limit $g \to \text{const}$ on the level of equivalence classes of relative S matrices, where the equivalence relation is given by unitary conjugation. While this construction as well as the verification of the Haag--Kastler axioms is by now well-known~\cite{fredenhagenrejzner2015}, it is somewhat scattered throughout the literature, such that we give here a self-contained exhibition.

We start from the functional equation~\cite{brunettifredenhagen2000}
\begin{equation}
\label{eq:smatrix_factorization}
\hat{S}(g+f+h) = \hat{S}(g+f) \hat{S}(g)^{-1} \hat{S}(g+h) \eqend{,}
\end{equation}
which should hold whenever $f$ is supported in the future of $h$ such that $J^+(\supp f) \cap J^-(\supp h) = \emptyset$. Multiplying by $\hat{S}(g)^{-1}$ from the left, we can also write it as~\cite{bogoljubowschirkow1955}
\begin{equation}
\label{eq:relative_smatrix_factorization}
\hat{S}_g(f+h) = \hat{S}_g(f) \hat{S}_g(h) \eqend{,}
\end{equation}
i.e., the relative S matrices also factorize in a causal way. In perturbation theory, this relation follows from the causal factorization condition~\eqref{eq:time_ordered_product_causalfactorization} for any relativistic QFT order by order, see for example~\cite[Thm.~2]{froeb2018}. Since we have shown in section~\ref{sec:smatrix_series} that the perturbative series for $\hat{S}$ converges strongly, the Cauchy product of the series on the right-hand side of~\eqref{eq:smatrix_factorization} converges to the product of the individual series, and therefore the relation~\eqref{eq:smatrix_factorization} indeed holds for the full $\hat{S}$.

\begin{figure}[ht]
\centering
\includegraphics{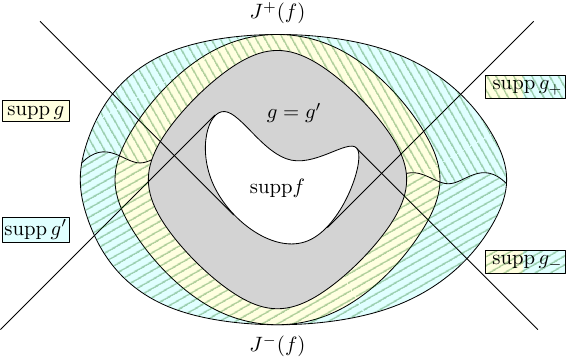}
\caption{Decomposition of the difference between two adiabatic cutoff functions $g$ and $g'$, which coincide in a neighborhood of $\supp f$. We can decompose the difference $g - g'$ into $g_+ + g_-$, where the cut is made at spacelike separation from $\supp f$ such that $J^+(\supp g_+) \cap J^-(\supp f) = \emptyset$ and $J^-(\supp g_-) \cap J^+(\supp f) = \emptyset$.}
\label{fig:decomposition}
\end{figure}
For the subsequent derivation, we essentially follow~\cite[Prop.~8.1]{brunettifredenhagen2000}. Consider two adiabatic cutoff functions $g$ and $g'$, which coincide in a neighborhood of the causal hull of $\supp f$. We can then make the decomposition $g' = g + g_+ + g_-$, where the functions $g_\pm$ are such that $J^+(\supp g_+) \cap J^-(\supp f) = \emptyset$ and $J^-(\supp g_-) \cap J^+(\supp f) = \emptyset$, see Fig.~\ref{fig:decomposition} for an illustration. From the relation~\eqref{eq:smatrix_factorization} with the replacements $g \to g + g_-$, $f \to g_+$ and $h \to f$, we obtain that
\begin{equation}
\hat{S}(g'+f) = \hat{S}(g') \hat{S}(g+g_-)^{-1} \hat{S}(g+g_-+f)
\end{equation}
and hence, multiplying with $\hat{S}(g')^{-1}$ from the left,
\begin{equation}
\label{eq:relative_smatrix_causalrelation1}
\hat{S}_{g'}(f) = \hat{S}(g')^{-1} \hat{S}(g'+f) = \hat{S}(g+g_-)^{-1} \hat{S}(g+g_-+f) = \hat{S}_{g+g_-}(f) \eqend{.}
\end{equation}
On the other hand, using relation~\eqref{eq:smatrix_factorization} with the replacement $h \to g_-$ shows that
\begin{equation}
\hat{S}(g+f+g_-) = \hat{S}(g+f) \hat{S}(g)^{-1} \hat{S}(g+g_-) \eqend{,}
\end{equation}
and hence
\begin{splitequation}
\label{eq:relative_smatrix_causalrelation2}
\hat{S}_{g+g_-}(f) &= \hat{S}(g+g_-)^{-1} \hat{S}(g+f) \hat{S}(g)^{-1} \hat{S}(g+g_-) \\
&= \left[ \hat{S}(g+g_-)^{-1} \hat{S}(g) \right] \hat{S}(g)^{-1} \hat{S}(g+f) \left[ \hat{S}(g+g_-)^{-1} \hat{S}(g) \right]^{-1} \eqend{.}
\end{splitequation}
Combining the identities~\eqref{eq:relative_smatrix_causalrelation1} and~\eqref{eq:relative_smatrix_causalrelation2}, it follows that
\begin{equation}
\label{eq:relative_smatrix_conjugation}
\hat{S}_{g'}(f) = U(g,g') \hat{S}_g(f) U^{-1}(g,g')
\end{equation}
with the unitaries
\begin{equation}
\label{eq:relative_smatrix_conjugation_unitaries}
U(g,g') = \hat{S}(g+g_-)^{-1} \hat{S}(g) = \hat{S}_g(g_-)^{-1} \eqend{.}
\end{equation}
That is, a change of the adiabatic cutoff function $g \to g'$ outside the causal hull of $f$ only results in a unitary conjugation of the relative S matrix.

To define the equivalence classes of relative S matrices, we first define equivalence classes of adiabatic cutoff functions by
\begin{equation}
[G]_M \equiv \{ g\colon g = G = \text{const. on the region } (\tau,\theta)\colon \tau \in M \} \eqend{,}
\end{equation}
i.e., we consider all test functions $g$ which are constant on the smallest time slice that contains the region $M$. Since global de~Sitter spacetime has compact spatial sections and the bound~\eqref{eq:smatrix_cg_def} is finite for such test functions $g$, this is a suitable choice. However, it does not have any influence on the subsequent arguments, and other choices for the equivalence classes are possible. Given a region $M$ and a test function $f$ such that $\supp f \subset M$, we then define equivalence classes of relative S matrices
\begin{equation}
\label{eq:relative_smatrix_equivalence_def}
\mathcal{S}_{G,M}(f) \equiv \left[ \hat{S}_g(f) \right]_{[G]_M} = \left\{ \hat{S}_g(f)\colon g \in [G]_M \right\} / \sim \eqend{,}
\end{equation}
with the equivalence relation $\sim$ being given by unitary conjugation. That is, we have
\begin{equation}
\hat{S}_g(f) \sim \hat{S}_{g'}(f) \quad \Leftrightarrow \quad \exists V\colon V^\dagger = V^{-1} \eqend{,} \ \hat{S}_{g'}(f) = V \hat{S}_g(f) V^{-1}
\end{equation}
which indeed holds for all adiabatic cutoffs $g \in [G]_M$ with $V = U(g,g')$~\eqref{eq:relative_smatrix_conjugation_unitaries} according to~\eqref{eq:relative_smatrix_conjugation}, such that $\mathcal{S}_{G,M}(f)$ is well defined. One easily checks that also their products are well defined, i.e., that two different representatives of a product are equivalent, such that
\begin{equation}
\mathcal{S}_{g'}(f) \mathcal{S}_{g'}(f') \sim \mathcal{S}_g(f) \mathcal{S}_g(f')
\end{equation}
holds for $\supp f, \supp f' \subset M$ and $g, g' \in [G]_M$.

The Haag--Kastler net (in the algebraic adiabatic limit) is then defined by the local algebras $\mathfrak{A}_G(M)$, which are the algebras generated by the $\mathcal{S}_{G,M}(f)$. It remains to verify the axioms of isotony, locality and (de~Sitter) covariance.

Given two regions $M_1 \subset M_2$, we define the embeddings
\begin{equation}
i_{M_2 M_1} \colon \mathfrak{A}_G(M_1) \to \mathfrak{A}_G(M_2) \eqend{,} \quad \mathfrak{A}_G(M_1) \ni \mathcal{S}_{G,M_1}(f) \mapsto \mathcal{S}_{G,M_2}(f) \in \mathfrak{A}_G(M_2) \eqend{.}
\end{equation}
This embedding of local algebras is well-defined on the level of representatives which are such that $g \in [G]_{M_2}$ (since then also $g \in [G]_{M_1}$), and such a representative can always be found because $M_1 \subset M_2$. The verification of the isotony axiom $i_{M_3 M_2} \circ i_{M_2 M_1} = i_{M_3 M_1}$ for $M_1 \subset M_2 \subset M_3$ is then an easy computation, inserting the various definitions for some $g \in [G]_{M_3}$. For the locality axiom, let us consider two test functions $f$ and $h$ whose supports are spacelike separated, i.e., $J^\pm(\supp f) \cap J^\mp(\supp h) = \emptyset$. Then the causal factorization of relative S matrices~\eqref{eq:relative_smatrix_factorization} holds also with $f$ and $h$ exchanged, such that
\begin{equation}
\hat{S}_g(f) \hat{S}_g(h) = \hat{S}_g(h) \hat{S}_g(f)
\end{equation}
for all $g$. It follows that this relation also holds for the equivalence classes $\mathcal{S}_{G,M}(f)$~\eqref{eq:relative_smatrix_equivalence_def}, i.e., the commutator $[ S_{G,M}(f), S_{G,M}(h) ] = 0$ vanishes. Choosing spacelike separated spacetime regions $M_1$ and $M_2$ with $\supp f \subset M_1$ and $\supp h \subset M_2$, and a bigger region $M$ such that $M_1, M_2 \subset M$, a short computation reveals the locality condition for the net:
\begin{equation}
\left[ i_{M M_1}\left( \mathfrak{A}_G(M_1) \right), i_{M M_2}\left( \mathfrak{A}_G(M_2) \right) \right] = \{ 0 \} \eqend{.}
\end{equation}

Instead of Poincar{\'e} covariance, we have to show de~Sitter covariance of the net. This is a bit more involved, and we first note that from the transformation~\eqref{eq:smatrix_symmetry_transform} of the S matrix we obtain the transformation
\begin{equation}
\label{eq:relative_smatrix_transform}
U_{abc} \, \hat{S}_g(f) U_{abc}^\dagger = \hat{S}_{g_{abc}}(f_{abc})
\end{equation}
of the relative S matrices~\eqref{eq:relative_smatrix_def}. We recall that $f_{abc}(x) = f(x_{abc})$, where the transformed coordinates $x_{abc}$ are defined by~\eqref{eq:appendix_ds_coordinates}. Given a region $M$, we define the transformed region $M_{abc}$ by
\begin{equation}
M_{abc} \equiv \{ x_{abc}\colon x \in M \} \eqend{,}
\end{equation}
and it follows that
\begin{equation}
\label{eq:transformed_testfunction_support}
\supp g \subset M \quad\Rightarrow \supp g_{abc} \subset M_{abc} \eqend{.}
\end{equation}
Covariance of the net then means that there exists an isometric map $\alpha_{abc}^M\colon \mathfrak{A}_G(M) \to \mathfrak{A}_G(M_{abc})$ which intertwines the algebras for $M_1 \subset M_2$:
\begin{equation}
\alpha_{abc}^{M_2} \circ i_{M_2,M_1} = i_{M_{2,abc}, M_{1,abc}} \circ \alpha_{abc}^{M_1} \eqend{.}
\end{equation}
That is, if we first transform an element of $\mathfrak{A}_G(M_1)$, which after the transformation is an element of $\mathfrak{A}_G(M_{1,abc})$, and then embed into the algebra $\mathfrak{A}_G(M_{2,abc})$ we have the same result as if we first embed the element that we start with into $\mathfrak{A}_G(M_2)$, and then transform to obtain an element of $\mathfrak{A}_G(M_{2,abc})$.

We tentatively define the maps $\alpha$ by
\begin{splitequation}
\label{eq:relative_smatrix_covariance_alpha_def}
\alpha_{abc}^M&\colon \mathfrak{A}_G(M) \to \mathfrak{A}_G(M_{abc}) \eqend{,} \\
\mathfrak{A}_G(M) \ni S_{G,M}(f) &\mapsto U_{abc} \, S_{G,M}(f) U_{abc}^\dagger = S_{G,M_{abc}}(f_{abc}) \in \mathfrak{A}_G(M_{abc}) \eqend{,}
\end{splitequation}
which are clearly isometric. Again we have to check that they are well-defined: we choose $g \in [G]_M$ and $f$ with $\supp f \subset M$ such that the relative S matrix $\hat{S}_g(f)$ is a representative of $S_{G,M}(f)$. By the support properties~\eqref{eq:transformed_testfunction_support} of the transformed test functions, it follows that the transformed relative S matrix $\hat{S}_{g_{abc}}(f_{abc})$~\eqref{eq:relative_smatrix_transform} is also a representative of $S_{G,M_{abc}}(f_{abc})$. Choosing a different representative $g' \in [G]_M$, we compute
\begin{equation}
U_{abc} \hat{S}_{g'}(f) U_{abc}^\dagger = \hat{S}_{g'_{abc}}(f_{abc}) = U(g_{abc},g'_{abc}) \hat{S}_{g_{abc}}(f_{abc}) U(g_{abc},g'_{abc})^{-1}
\end{equation}
with the unitaries $U(g,g')$ defined by~\eqref{eq:relative_smatrix_conjugation_unitaries}. Namely, if both $g,g' \in [G]_M$, then $g_{abc},g'_{abc} \in [G]_{M_{abc}}$, and therefore the unitary conjugation~\eqref{eq:relative_smatrix_conjugation} holds. It follows that also $\hat{S}_{g'_{abc}}(f_{abc})$ is a representative of $S_{G,M_{abc}}(f_{abc})$, and the map $\alpha_{abc}^M$~\eqref{eq:relative_smatrix_covariance_alpha_def} is indeed well-defined. To verify the intertwining relation, we then simply compute for $\supp f \subset M_1 \subset M_2$ that
\begin{equation}
( \alpha_{abc}^{M_2} \circ i_{M_2,M_1} ) S_{G,M_1}(f) = \alpha_{abc}^{M_2} S_{G,M_2}(f) = S_{G,M_{2,abc}}(f_{abc}) \in \mathfrak{A}_G(M_{2,abc})
\end{equation}
and that
\begin{equation}
( i_{M_{2,abc}, M_{1,abc}} \circ \alpha_{abc}^{M_1} ) S_{G,M_1}(f) = i_{M_{2,abc}, M_{1,abc}} S_{G,M_{1,abc}}(f_{abc}) = S_{G,M_{2,abc}}(f_{abc}) \eqend{,}
\end{equation}
and both expressions agree as required.

\section{Interacting observables}
\label{sec:observables}

Apart from the relative S matrices, also other interacting observables exist in the Sine--Gordon model. These are constructed using the Bogoliubov formula
\begin{equation}
\label{eq:bogoliubov}
\left[ \hat{\op}(x) \right]_\text{int} \equiv \hat{S}(g)^{-1} \mathcal{T}\left[ \op(x) \otimes \mathe_\otimes^{\mathi S_\text{int}} \right] \eqend{,}
\end{equation}
in analogy to the relative S matrices~\eqref{eq:relative_smatrix_def}, taking into account the definition of the S matrix~\eqref{eq:smatrix_def}. In particular, we are interested in the interacting vertex operators $\left[ \hat{V}_{\pm \beta} \right]_\text{int}$ and the interacting field $\hat{\phi}_\text{int}$ itself. Another important observable is the stress tensor $T_{\mu\nu}$, whose construction is however more complicated~\cite{froebcadamuro2022a}. Again, in general the Bogoliubov formula~\eqref{eq:bogoliubov} needs to be understood as a formal power series in $g$, but we will show that the series is convergent for the vertex operators and the field itself if $\beta^2 < 4 \pi$.

\subsection{Convergent perturbative series for the interacting observables}
\label{sec:observables_series}

The proof of the convergence of the perturbation expansion of the Bogoliubov formula~\eqref{eq:bogoliubov} proceeds in complete analogy with the case of the S matrix in section~\ref{sec:smatrix_series}. We smear the observable with a test function $h$, and obtain the series expansion $\big[ \hat{\op}(h) \big]_\text{int} = \sum_{k=0}^\infty \big[ \hat{\op}(h) \big]_\text{int,k}$ with
\begin{splitequation}
\label{eq:observable_series_expansion}
\left[ \hat{\op}(h) \right]_\text{int,k} \equiv \sum_{\ell=0}^k \frac{(-\mathi)^\ell \mathi^{k-\ell}}{\ell! (k-\ell)!} \overline{\mathcal{T}}\left[ S_\text{int}^{\otimes \ell} \right] \mathcal{T}\left[ \op(h) \otimes S_\text{int}^{\otimes (k-\ell)} \right] \eqend{,}
\end{splitequation}
where we used that the series expansion of the inverse $\hat{S}(g)^{-1}$ involves anti-time-ordered products $\overline{\mathcal{T}}$.

Again we show that the series $\sum_{k=0}^\infty \big[ \hat{\op}(h) \big]_\text{int,k}$ converges strongly on the dense domain $D$, which is the span of the vectors $\mathe^{\mathi \hat{\phi}(f)} \ket{0_\alpha}$. For this, we have to bound the norm
\begin{splitequation}
\label{eq:observable_series_expansion_norm}
\norm{ \left[ \hat{\op}(h) \right]_\text{int,k} \mathe^{\mathi \hat{\phi}(f)} \ket{0_\alpha} } &\leq \sum_{\ell=0}^k \frac{1}{\ell! (k-\ell)!} \sum_{\sigma_1, \ldots, \sigma_k = \pm 1} \Bigg\lVert \int\dotsi\int \overline{\mathcal{T}}\left[ \bigotimes_{i=1}^\ell \mathe^{\mathi \sigma_i \beta \phi(\tau_i,\theta_i)} \right] \\
&\quad\times \mathcal{T}\left[ \op(h) \otimes \bigotimes_{i=\ell+1}^k \mathe^{\mathi \sigma_i \beta \phi(\tau_i,\theta_i)} \right] \prod_{i=1}^k \frac{g(\tau_i,\theta_i)}{H^2 \sin^2(\tau_i)} \total \tau_i \total \theta_i \, \mathe^{\mathi \hat{\phi}(f)} \ket{0_\alpha} \Bigg\rVert \eqend{,}
\end{splitequation}
which is done analogously to the previous computation. To avoid repetition, we thus only list the differences. First of all, we note that for vertex operators $\op(h) = V_{\pm \beta}(h) \equiv \int V_{\pm \beta}(x) h(x) \sqrt{-g} \total^2 x$ we obtain exactly the same bounds~\eqref{eq:smatrix_term_normsquared_erg} as before, but with one of the adiabatic cutoff functions $g$ replaced by $h$ and with $(k+1)!$ instead of $k!$. Summing over $\ell$, it follows that
\begin{equation}
\norm{ \left[ V_{\pm \beta}(h) \right]_\text{int,k} \mathe^{\mathi \hat{\phi}(f)} \ket{0_\alpha} } \leq (k+1) [(k+1)!]^{- \frac{4 \pi - \beta^2}{8 \pi}} C(h) [ 4 C(g) ]^k \eqend{,}
\end{equation}
where one factor $2^k$ comes again from the sum over the $\sigma_i$. This is summable in $k$ under the same conditions as for the S matrix, namely $\beta^2 < 4 \pi$ and $C(g) < \infty$, and it follows that the sum $\sum_{k=0}^\infty \big[ V_{\pm \beta}(h) \big]_\text{int,k}$ converges strongly on the dense domain $D$, such that the interacting vertex operator $\big[ V_{\pm \beta}(h) \big]_\text{int}$ is a well-defined operator on this domain.

For the field operator itself, we instead have to derive the required bounds anew, for which we can use the formula~\eqref{eq:vertex_phi_phi_weyl_correlator_neutral}. For the smeared field $\hat{\phi}(h)$, we choose a compactly supported test function $h$ which satisfies $\int h(x) \sqrt{-g} \total^2 x = 0$, since then we have
\begin{equation}
\hat{\phi}(h) = \int \hat{\phi}(x) h(x) \sqrt{-g} \total^2 x = \int \hat{\phi}(x) \partial_\mu \left[ h^\mu(x) \sqrt{-g} \right] \total^2 x = - \int \partial_\mu \hat{\phi}(x) h^\mu(x) \sqrt{-g} \total^2 x
\end{equation}
with another compactly supported test function $h^\mu$,\footnote{This is ensured by a version of the Poincar{\'e} lemma for functions with compact support, see for example~\cite[App.~B and~C]{froebhackhiguchi2017} for a proof.} such that only derivatives of $\hat{\phi}$ enter. In this case, in the result~\eqref{eq:vertex_phi_phi_weyl_correlator_neutral} we can take the limit $\alpha \to \infty$ in which it simplifies, and we obtain
\begin{splitequation}
\label{eq:field_correlator_neutral_derivative_smared}
&\lim_{\alpha \to \infty} \bra{0_\alpha} \mathe^{- \mathi \hat{\phi}(f)} \hat{\phi}(h) V_{\gamma_1}(\tau_1,\theta_1) \cdots V_{\gamma_n}(\tau_n,\theta_n) \hat{\phi}(h) \, \mathe^{\mathi \hat{\phi}(f)} \ket{0_\alpha} \Big\rvert_{\raisebox{0.2em}{\scriptsize $\sum_{j=1}^n \gamma_j = 0$}} \\
&\quad= \lim_{\alpha \to \infty} \bra{0_\alpha} \mathe^{- \mathi \hat{\phi}(f)} V_{\gamma_1}(\tau_1,\theta_1) \cdots V_{\gamma_n}(\tau_n,\theta_n) \, \mathe^{\mathi \hat{\phi}(f)} \ket{0_\alpha} \Big\rvert_{\raisebox{0.2em}{\scriptsize $\sum_{j=1}^n \gamma_j = 0$}} \\
&\qquad\times \iint \Bigg[ \bigg[ G^+_\alpha(f,\tau',\theta') - G^+_\alpha(\tau',\theta',f) - \frac{\mathi}{4 \pi} \sum_{k=1}^n \gamma_k \ln\abs{ 2 \cos(\tau'-\tau_k) - 2 \cos(\theta'-\theta_k) } \\
&\hspace{8em}+ \frac{1}{4} \sum_{k=1}^n \gamma_k \Theta\left[ \cos(\theta'-\theta_k) - \cos(\tau'-\tau_k) \right] \sgn \sin(\tau'-\tau_k) \bigg] \\
&\qquad\qquad\ \times \bigg[ G^+_\alpha(f,\tau'',\theta'') - G^+_\alpha(\tau'',\theta'',f) - \frac{\mathi}{4 \pi} \sum_{k=1}^n \gamma_k \ln\abs{ 2 \cos(\tau_k-\tau'') - 2 \cos(\theta_k-\theta'') } \\
&\hspace{8em}- \frac{1}{4} \sum_{k=1}^n \gamma_k \Theta\left[ \cos(\theta''-\theta_k) - \cos(\tau''-\tau_k) \right] \sgn \sin(\tau''-\tau_k) \bigg] \\
&\qquad\qquad- \frac{\mathi}{4} \Theta\left[ \cos(\theta'-\theta'') - \cos(\tau'-\tau'') \right] \sgn \sin(\tau'-\tau'') \\
&\qquad\qquad- \frac{1}{4 \pi} \ln\abs{ 2 \cos(\tau'-\tau'') - 2 \cos(\theta'-\theta'') } \Bigg] \frac{h(\tau',\theta')}{H^2 \sin^2(\tau')} \total \tau' \total \theta' \frac{h(\tau'',\theta'')}{H^2 \sin^2(\tau'')} \total \tau'' \total \theta'' \eqend{,}
\end{splitequation}
where we used~\eqref{eq:free_scalar_2pf_log} to perform the limit $\epsilon \to 0^+$. The remaining correlation function is given by~\eqref{eq:vertex_weyl_correlator} and~\eqref{eq:vertex_correlator_neutral}, and as before the corresponding result involving time-ordered or anti-time-ordered products is obtained by the simple replacements $\tau_j - \tau_k \to \abs{\tau_j - \tau_k}$ or $\tau_j - \tau_k \to - \abs{\tau_j - \tau_k}$. However, since we only consider the case $- \gamma_j \gamma_k < 4 \pi$, all singularities are integrable such that we can take the limit $\epsilon \to 0^+$ everywhere, and then we always obtain the same expression.

Namely, we compute
\begin{splitequation}
&\norm{ \int\dotsi\int \overline{\mathcal{T}}\left[ \bigotimes_{i=1}^\ell \mathe^{\mathi \sigma_i \beta \phi(\tau_i,\theta_i)} \right] \mathcal{T}\left[ \phi(h) \otimes \bigotimes_{i=\ell+1}^k \mathe^{\mathi \sigma_i \beta \phi(\tau_i,\theta_i)} \right] \prod_{i=1}^k \frac{g(\tau_i,\theta_i)}{H^2 \sin^2(\tau_i)} \total \tau_i \total \theta_i \, \mathe^{\mathi \hat{\phi}(f)} \ket{0_\alpha} }^2 \\
&= \int\dotsi\int \bra{0_\alpha} \mathe^{- \mathi \hat{\phi}(f)} \, \overline{\mathcal{T}}\left[ \phi(h) \otimes \! \bigotimes_{i=\ell+1}^k \mathe^{- \mathi \sigma_i \beta \phi(\tau_i,\theta_i)} \right] \mathcal{T}\left[ \bigotimes_{i=1}^\ell \mathe^{- \mathi \sigma_i \beta \phi(\tau_i,\theta_i)} \right] \\
&\quad\times \overline{\mathcal{T}}\left[ \bigotimes_{i=k+1}^{k+\ell} \mathe^{\mathi \sigma_{i-k} \beta \phi(\tau_i,\theta_i)} \right] \mathcal{T}\left[ \phi(h) \otimes \!\!\! \bigotimes_{i=k+\ell+1}^{2k} \!\! \mathe^{\mathi \sigma_{i-k} \beta \phi(\tau_i,\theta_i)} \right] \mathe^{\mathi \hat{\phi}(f)} \ket{0_\alpha} \prod_{i=1}^{2k} \frac{g(\tau_i,\theta_i)}{H^2 \sin^2(\tau_i)} \total \tau_i \total \theta_i \eqend{,}
\end{splitequation}
where we used that the adjoint of a time-ordered product is an anti-time-ordered product. Since we consider the case $\beta^2 < 4 \pi$, the time-ordered and anti-time-ordered products can be defined everywhere by continuity, and the correlator is in fact independent of the time-ordering (up to a possible phase, which vanishes in our case). The resulting correlator satisfies the neutrality condition, and we can thus use the result~\eqref{eq:field_correlator_neutral_derivative_smared} to obtain (in the limit $\alpha \to \infty$)
\begin{splitequation}
\label{eq:field_correlator_norm_limit}
&\norm{ \int\dotsi\int \overline{\mathcal{T}}\left[ \bigotimes_{i=1}^\ell \mathe^{\mathi \sigma_i \beta \phi(\tau_i,\theta_i)} \right] \mathcal{T}\left[ \phi(h) \otimes \bigotimes_{i=\ell+1}^k \mathe^{\mathi \sigma_i \beta \phi(\tau_i,\theta_i)} \right] \prod_{i=1}^k \frac{g(\tau_i,\theta_i)}{H^2 \sin^2(\tau_i)} \total \tau_i \total \theta_i \, \mathe^{\mathi \hat{\phi}(f)} \ket{0_\alpha} }^2 \\
&\quad\to \int\dotsi\int \lim_{\alpha \to \infty} \bra{0_\alpha} \mathe^{- \mathi \hat{\phi}(f)} \prod_{i=1}^k V_{-\sigma_i\beta}(\tau_i,\theta_i) \prod_{i=k+1}^{2k} V_{\sigma_{i-k} \beta}(\tau_i,\theta_i) \, \mathe^{\mathi \hat{\phi}(f)} \ket{0_\alpha} \\
&\qquad\times \iint \Bigg[ K_+\left( f,\tau',\theta',\{\sigma_j\},\{\tau_j,\theta_j\} \right) K_-\left( f,\tau'',\theta'',\{\sigma_j\},\{\tau_j,\theta_j\} \right) \\
&\qquad\qquad- \frac{\mathi}{4} \Theta\left[ \cos(\theta'-\theta'') - \cos(\tau'-\tau'') \right] \sgn \sin(\tau'-\tau'') \\
&\qquad\qquad- \frac{1}{4 \pi} \ln\abs{ 2 \cos(\tau'-\tau'') - 2 \cos(\theta'-\theta'') } \Bigg] \frac{h(\tau',\theta')}{H^2 \sin^2(\tau')} \total \tau' \total \theta' \frac{h(\tau'',\theta'')}{H^2 \sin^2(\tau'')} \total \tau'' \total \theta'' \\
&\qquad\times \prod_{i=1}^{2k} \frac{g(\tau_i,\theta_i)}{H^2 \sin^2(\tau_i)} \total \tau_i \total \theta_i \raisetag{2em}
\end{splitequation}
with the functions
\begin{splitequation}
\label{eq:field_correlator_kdef}
&K_\pm\left( f,\tau',\theta',\{\sigma_j\},\{\tau_j,\theta_j\} \right) \\
&\ \equiv - \frac{1}{2} \iint \Theta\left[ \cos(\theta-\theta') - \cos(\tau-\tau') \right] \sgn \sin(\tau-\tau') \frac{f(\tau,\theta)}{H^2 \sin^2(\tau)} \total \tau \total \theta \\
&\quad+ \frac{\mathi \beta}{4 \pi} \sum_{j=1}^k \sigma_j \bigg[ \ln\abs{ 2 \cos(\tau'-\tau_j) - 2 \cos(\theta'-\theta_j) } - \ln\abs{ 2 \cos(\tau'-\tau_{k+j}) - 2 \cos(\theta'-\theta_{k+j}) } \\
&\hspace{6em}\pm \mathi \pi \Theta\left[ \cos(\theta'-\theta_j) - \cos(\tau'-\tau_j) \right] \sgn \sin(\tau'-\tau_j) \\
&\hspace{6em}\mp \mathi \pi \Theta\left[ \cos(\theta'-\theta_{k+j}) - \cos(\tau'-\tau_{k+j}) \right] \sgn \sin(\tau'-\tau_{k+j}) \bigg] \eqend{.} \raisetag{2em}
\end{splitequation}
We first bound the integrals over $\tau'$, $\theta'$, $\tau''$ and $\theta''$, in such a way that the bound is independent of the remaining points $(\tau_i,\theta_i)$. This will allow us then to use the previous bounds for the remaining integrals.

Clearly we have $\abs{ \Theta(x) \sgn(y) } \leq 1$, and can thus bound
\begin{splitequation}
\abs{ K_\pm\left( f,\tau',\theta',\{\sigma_j\},\{\tau_j,\theta_j\} \right) } &\leq \frac{1}{2} \norm{ \frac{f(\tau,\theta)}{H^2 \sin^2(\tau)} }_1 + k \frac{\beta}{2} \\
&\quad+ \frac{\beta}{4 \pi} \sum_{j=1}^{2k} \Big\lvert \ln\abs{ 2 \cos(\tau'-\tau_j) - 2 \cos(\theta'-\theta_j) } \Big\rvert \eqend{.}
\end{splitequation}
To integrate this bound with the test function $h$ over $\tau'$ and $\theta'$, we first pass to coordinates $u$ and $v$ according to~\eqref{eq:light_cone_coordinates_def} and then use the identities~\eqref{eq:cosine_identities}, such that
\begin{splitequation}
\abs{ K_\pm\left( f,\tau',\theta',\{\sigma_j\},\{\tau_j,\theta_j\} \right) } &\leq \frac{1}{2} \norm{ \frac{f(\tau,\theta)}{H^2 \sin^2(\tau)} }_1 + k \frac{\beta}{2} \\
&\quad+ \frac{\beta}{4 \pi} \sum_{j=1}^{2k} \left( \abs{ \ln\abs{ \mathe^{\mathi u'} - \mathe^{\mathi u_j} } } + \abs{ \ln\abs{ \mathe^{\mathi v'} - \mathe^{\mathi v_j} } } \right) \eqend{.}
\end{splitequation}
Using the Hölder inequality in the limit $p \to 1$, we then compute
\begin{splitequation}
\label{eq:field_correlator_bound_k_int}
&\abs{ \int K_+\left( f,\tau',\theta',\{\sigma_j\},\{\tau_j,\theta_j\} \right) \frac{h(\tau',\theta')}{H^2 \sin^2(\tau')} \total \tau' \total \theta' } \\
&\quad\leq \frac{1}{2} \left[ \norm{ \frac{f(\tau,\theta)}{H^2 \sin^2(\tau)} }_1 + k \beta \right] \norm{ \frac{h(\tau,\theta)}{H^2 \sin^2(\tau)} }_1 \\
&\qquad+ \frac{\beta}{8 \pi} \norm{ \frac{h(\tau,\theta)}{H^2 \sin^2(\tau)} }_\infty \sum_{j=1}^{2k} \iint \left( \abs{ \ln\abs{ \mathe^{\mathi u'} - \mathe^{\mathi u_j} } } + \abs{ \ln\abs{ \mathe^{\mathi v'} - \mathe^{\mathi v_j} } } \right) \total u' \total v' \eqend{,}
\end{splitequation}
and furthermore
\begin{splitequation}
\iint \abs{ \ln\abs{ \mathe^{\mathi u'} - \mathe^{\mathi u_j} } } \total u' \total v' &= 3 \pi \int_{-2\pi-u_j}^{\pi-u_j} \abs{ \ln\abs{ \mathe^{\mathi (u'+u_j)} - \mathe^{\mathi u_j} } } \total u' \leq 3 \pi^2 \int_{-3}^3 \abs{ \ln\abs{ \mathe^{\mathi \pi s} - 1 } } \total s \\
&= 9 \pi^2 \left[ \int_1^\frac{5}{3} \ln\left[ 2 - 2 \cos(\pi s) \right] \total s + \int_{- \frac{1}{3}}^0 \ln\left[ \frac{1}{2 - 2 \cos(\pi s)} \right] \total s \right] \eqend{,}
\end{splitequation}
where we took into account the range of $u$~\eqref{eq:light_cone_coordinates_def} as well as the periodicity of the integrand, and used the identity~\eqref{eq:cosine_identity_2}. The first integral can be bounded by bounding the integrand by its maximal value at $s = 1$, while for the second we use that the logarithm is a monotonely growing function in that interval, such that we can bound the integral by bounding the cosine~\eqref{eq:appendix_cosine_bound}. It follows that
\begin{splitequation}
\iint \abs{ \ln\abs{ \mathe^{\mathi u'} - \mathe^{\mathi u_j} } } \total u' \total v' &\leq 9 \pi^2 \left[ \frac{4}{3} \ln 2 + \int_{-\frac{1}{3}}^0 \ln\left[ \frac{1}{8 s^2} \right] \total s \right] \\
&= 3 \pi^2 \big( \ln 2 + 2 + 2 \ln 3 \big) \leq 15 \pi^2 \eqend{,}
\end{splitequation}
and the same bound for the term involving $v'$. The bound~\eqref{eq:field_correlator_bound_k_int} thus reduces to
\begin{splitequation}
\label{eq:field_correlator_bound_k_int_2}
&\abs{ \int K_+\left( f,\tau',\theta',\{\sigma_j\},\{\tau_j,\theta_j\} \right) \frac{h(\tau',\theta')}{H^2 \sin^2(\tau')} \total \tau' \total \theta' } \\
&\quad\leq \frac{1}{2} \norm{ \frac{f(\tau,\theta)}{H^2 \sin^2(\tau)} }_1 \norm{ \frac{h(\tau,\theta)}{H^2 \sin^2(\tau)} }_1 + \frac{k \beta}{2} \left[ \norm{ \frac{h(\tau,\theta)}{H^2 \sin^2(\tau)} }_1 + 15 \pi \norm{ \frac{h(\tau,\theta)}{H^2 \sin^2(\tau)} }_\infty \right] \eqend{.}
\end{splitequation}

The other terms in the limit~\eqref{eq:field_correlator_norm_limit} are bounded similarly: we obtain
\begin{splitequation}
&\bigg\lvert \iint \frac{\mathi}{4} \Theta\left[ \cos(\theta'-\theta'') - \cos(\tau'-\tau'') \right] \sgn \sin(\tau'-\tau'') \frac{h(\tau',\theta')}{H^2 \sin^2(\tau')} \total \tau' \total \theta' \\
&\qquad\times \frac{h(\tau'',\theta'')}{H^2 \sin^2(\tau'')} \total \tau'' \total \theta'' \bigg\rvert \leq \frac{1}{4} \norm{ \frac{h(\tau,\theta)}{H^2 \sin^2(\tau)} }_1^2
\end{splitequation}
and
\begin{splitequation}
&\abs{ \iint \frac{1}{4 \pi} \ln\abs{ 2 \cos(\tau'-\tau'') - 2 \cos(\theta'-\theta'') } \frac{h(\tau',\theta')}{H^2 \sin^2(\tau')} \total \tau' \total \theta' \frac{h(\tau'',\theta'')}{H^2 \sin^2(\tau'')} \total \tau'' \total \theta'' } \\
&\quad\leq \frac{135}{2} \pi^3 \norm{ \frac{h(\tau,\theta)}{H^2 \sin^2(\tau)} }_\infty^2 \eqend{,}
\end{splitequation}
and as required all bounds are independent of the other points $(\tau_i,\theta_i)$. For the remaining integrals in~\eqref{eq:field_correlator_norm_limit} we can thus follow the same steps as for the S matrix bounds~\eqref{eq:smatrix_term_normsquared_erg}, in particular use the results~\eqref{eq:vertex_weyl_correlator} and~\eqref{eq:vertex_correlator_neutral}, and finally obtain
\begin{splitequation}
\label{eq:field_correlator_norm_limit_2}
&\lim_{\alpha \to \infty} \norm{ \int\dotsi\!\int \overline{\mathcal{T}}\left[ \bigotimes_{i=1}^\ell \mathe^{\mathi \sigma_i \beta \phi(\tau_i,\theta_i)} \right] \mathcal{T}\left[ \phi(h) \otimes \!\! \bigotimes_{i=\ell+1}^k \mathe^{\mathi \sigma_i \beta \phi(\tau_i,\theta_i)} \right] \prod_{i=1}^k \frac{g(\tau_i,\theta_i)}{H^2 \sin^2(\tau_i)} \total \tau_i \total \theta_i \, \mathe^{\mathi \hat{\phi}(f)} \ket{0_\alpha} }^2 \\
&\quad\leq (k!)^{1 + \frac{\beta^2}{4 \pi}} C(g)^{2k} \Big[ \tilde{C}_{(0)}(f,h) + 2 k \tilde{C}_{(1)}(f,h) + k^2 \tilde{C}_{(2)}(f,h) \Big] \\
&\quad\leq (k!)^{1 + \frac{\beta^2}{4 \pi}} C(g)^{2k} (k+1)^2 \max_{i \in \{1,2,3\}} \tilde{C}_{(i)}(f,h) \raisetag{2em}
\end{splitequation}
with $C(g)$ defined in~\eqref{eq:smatrix_cg_def}, and the constants
\begin{equations}[eq:smatrix_cfh_def]
\tilde{C}_{(0)}(f,h) &\equiv \frac{1}{4} \left( 1 + \norm{ \frac{f(\tau,\theta)}{H^2 \sin^2(\tau)} }_1^2 \right) \norm{ \frac{h(\tau,\theta)}{H^2 \sin^2(\tau)} }_1^2 + \frac{135}{2} \pi^3 \norm{ \frac{h(\tau,\theta)}{H^2 \sin^2(\tau)} }_\infty^2 \eqend{,} \\
\tilde{C}_{(1)}(f,h) &\equiv \beta \norm{ \frac{f(\tau,\theta)}{H^2 \sin^2(\tau)} }_1 \norm{ \frac{h(\tau,\theta)}{H^2 \sin^2(\tau)} }_1 \left[ \norm{ \frac{h(\tau,\theta)}{H^2 \sin^2(\tau)} }_1 + 15 \pi \norm{ \frac{h(\tau,\theta)}{H^2 \sin^2(\tau)} }_\infty \right] \eqend{,} \\
\tilde{C}_{(2)}(f,h) &\equiv \beta^2 \left[ \norm{ \frac{h(\tau,\theta)}{H^2 \sin^2(\tau)} }_1 + 15 \pi \norm{ \frac{h(\tau,\theta)}{H^2 \sin^2(\tau)} }_\infty \right]^2 \eqend{.}
\end{equations}
While we have considered the limit $\alpha \to \infty$ for simplicity, the bounds for finite $\alpha$ follow analoguously, with the only difference that the constants $\tilde{C}_{(i)}(f,h)$ then depend on $\alpha$.

For the norm~\eqref{eq:observable_series_expansion_norm}, we thus obtain the bound
\begin{equation}
\label{eq:interacting_field_norm_sum}
\norm{ \left[ \hat{\phi}(h) \right]_\text{int,k} \mathe^{\mathi \hat{\phi}(f)} \ket{0_\alpha} } \leq (k+1) (k!)^{- \frac{4 \pi - \beta^2}{8 \pi}} [ 4 C(g) ]^k \sqrt{ \max_{i \in \{1,2,3\}} \tilde{C}_{(i)}(f,h) } \eqend{,}
\end{equation}
where a factor $2^k$ again arises from the sum over the $\sigma_i = \pm 1$, and a factor $2^k/k!$ arises from the sum over $\ell$. This is summable in $k$ under the same conditions as for the S matrix and the interacting vertex operator, such that the sum $\sum_{k=0}^\infty \big[ \hat{\phi}(h) \big]_\text{int,k}$, with the test function $h$ satisfying the condition $\int h(x) \sqrt{-g} \total^2 x = 0$, converges strongly on the dense domain $D$ and the interacting field $\big[ \hat{\phi}(h) \big]_\text{int}$ is a well-defined operator on this domain. Note that while the overall magnitude of the bound~\eqref{eq:interacting_field_norm_sum} depends on $f$ and thus on the vector in $D$ on which the interacting field acts, the convergence rate does not, i.e., we have uniform convergence.

\subsection{Interacting field equation}
\label{sec:observables_eom}

By construction, the free quantized scalar field $\hat{\phi}$~\eqref{eq:free_scalar_fieldexpansion} fulfills the massless Klein--Gordon equation~\eqref{eq:free_scalar_eom}
\begin{equation}
\nabla^2 \hat{\phi}(\tau,\theta) = H^2 \sin^2(\tau) \, \partial^2 \hat{\phi}(\tau,\theta) = 0 \eqend{,}
\end{equation}
which is the equation of motion following from the free action~\eqref{eq:free_scalar_action}. We would like to verify that also the interacting field $\big[ \hat{\phi}(h) \big]_\text{int}$ satisfies the equation of motion following from the full Sine--Gordon action~\eqref{eq:sine_gordon_action}, which reads
\begin{equation}
\nabla^2 \phi - 2 g \beta \sin(\beta \phi) = 0 \eqend{.}
\end{equation}
Expressing the sine in terms of exponentials, interpreting those as vertex operators, and smearing the equation with a test function $h$, we thus want to show that
\begin{equation}
\label{eq:interacting_field_eom}
\big[ \hat{\phi}(\nabla^2 h) \big]_\text{int} - \mathi \beta \left[ V_{-\beta}(g h) \right]_\text{int} + \mathi \beta \left[ V_\beta(g h) \right]_\text{int} = 0 \eqend{.}
\end{equation}

The most efficient way to do this is to employ the relation
\begin{equation}
\label{eq:perturbative_agreement}
\mathcal{T}\left[ \phi(x) \otimes F^{\otimes n} \right] = \hat{\phi}(x) \, \mathcal{T}\left[ F^{\otimes n} \right] + \mathi n \int G^\text{ret}(y,x) \, \mathcal{T}\left[ F^{\otimes (n-1)} \otimes \frac{\delta F}{\delta \phi(y)} \right] \sqrt{-g} \total^2 y \eqend{,}
\end{equation}
which expresses a time-ordered product involving a single scalar field in terms of an ordinary product and the retarded propagator
\begin{equation}
G^\text{ret}(x,y) \equiv \Theta(x^0-y^0) \left[ G^+(x,y) - G^+(y,x) \right] \eqend{.}
\end{equation}
This relation expresses the principle of perturbative agreement~\cite{hollandswald2005,zahn2015,dragohackpinamonti2017}, namely that terms linear and quadratic in the fields can be shifted between free action and interaction without changing the quantum theory, in the linear case. In perturbation theory, it can always be fulfilled by a suitable choice of renormalization conditions for any relativistic QFT order by order, see for example~\cite[Thm.~9]{froeb2018}. Certainly it holds if no renormalization needs to be done and the time-ordered products can be extended to the diagonal by continuity, as in our case.

Since the retarded propagator fulfills $\nabla^2 G^\text{ret}(x,y) = \delta^2(x-y)/\sqrt{-g}$, smearing~\eqref{eq:perturbative_agreement} with $\nabla^2 h$ and integrating by parts it follows that
\begin{equation}
\label{eq:interacting_field_time_ordered_eom}
\mathcal{T}\left[ \phi(\nabla^2 h) \otimes F^{\otimes n} \right] = \mathi n \int \mathcal{T}\left[ F^{\otimes (n-1)} \otimes \frac{\delta F}{\delta \phi(x)} \right] h(x) \total^2 x \eqend{.}
\end{equation}
Taking now $F = \mathi S_\text{int} = 2 \mathi \int g(x) \cos[ \beta \phi(x) ] \sqrt{-g} \total^2 x$, we obtain
\begin{equation}
\frac{\delta F}{\delta \phi(x)} = - \beta g(x) \left[ \mathe^{\mathi \beta \phi(x)} - \mathe^{- \mathi \beta \phi(x)} \right] \sqrt{-g} \eqend{,}
\end{equation}
and dividing~\eqref{eq:interacting_field_time_ordered_eom} by $n!$ and summing over $n$ we obtain
\begin{equation}
\label{eq:interacting_field_time_ordered_eom2}
\mathcal{T}\left[ \phi(\nabla^2 h) \otimes \mathe_\otimes^{\mathi S_\text{int}} \right] = - \mathi \beta \int \mathcal{T}\left[ \left[ \mathe^{\mathi \beta \phi(x)} - \mathe^{- \mathi \beta \phi(x)} \right] \otimes \mathe_\otimes^{\mathi S_\text{int}} \right] g(x) h(x) \sqrt{-g} \total^2 x \eqend{.}
\end{equation}
Finally, multiplying this equation from the left by $\hat{S}(g)^{-1}$ and comparing with the Bogoliubov formula~\eqref{eq:bogoliubov} for interacting field operators, we obtain exactly the interacting field equation~\eqref{eq:interacting_field_eom}.

\section{Conclusion and outlook}

We have considered the Sine-Gordon model (with an adiabatic interaction cutoff) on two-dimensional de Sitter spacetime in the pAQFT framework, and showed the convergence of the S matrix, of the interacting vertex operator and of the interacting field operator on a dense domain in the Fock space of the free massless scalar field. For the free field, we considered a natural one-parameter family of ``vacuum'' states determined by the mode expansion of the massless field. However, the de Sitter symmetry is broken by those states for all values of the parameter, and only for derivatives of the field is restored in a limit. We then proceeded to construct the Haag--Kastler net of the interacting theory from the relative S-matrices, and showed that the Haag--Kastler axioms are fulfilled. As in previous works, the removal of the interaction cutoff is done on the level of the algebras, taking the so-called adiabatic algebraic limit, which is essential to verify the covariance axiom. Compared to Minkowski spacetime, one advantage of working on de Sitter spacetime is that the spatial sections are compact, such that there are no infrared issues even for the massless field, such that the field itself (and not only its derivative) is well-defined and we could work in the standard Fock space representation.

Future possible work includes the construction of other physically interesting observables such as the stress-energy tensor, analogously to the flat-space result~\cite{froebcadamuro2022a}, and the higher-order conserved currents~\cite{flume1975,yoon1976}. These should exist since the classical model admits an infinite number of conserved charges in flat spacetime~\cite{zakharovtakhtadzhyanfaddeev1974}, and one expects that integrability also holds in the quantum theory. While it has been shown recently that these currents are renormalizable at any order in perturbation theory~\cite{zanello2023}, the convergence of the perturbation series as well as the conservation of the renormalized currents remain open problems, even for Minkowski spacetime. Other avenues for future work include quantum energy inequalities for the Sine--Gordon model in de~Sitter spacetime, in analogy to the ones derived by us in flat spacetime~\cite{froebcadamuro2022b}. These are of fundamental importance for the stability of quantum field theory, both thermodynamically and regarding the structure of spacetime in the form of singularities; see for example Refs.~\cite{fewster2012,kontousanders2020,freivogelkontoukrommydas2022,fewsterkontou2022} for reviews and existing results.

It would also be interesting to extend our results to anti-de Sitter (AdS) spacetime, and to investigate the connection with the AdS/CFT correspondence. The Sine--Gordon model in AdS has been recently studied to low orders in perturbation theory~\cite{antunesetal2021}, and combining the methods presented here with theirs, it might be possible to obtain interesting all-order results. However, to apply the AdS/CFT dictionary, it is important to be able to take the adiabatic limit, where the interaction cutoff function $g$ becomes constant. Also this is an open problem, which is notoriously difficult and has not been resolved even in Minkowski spacetime\footnote{For the Euclidean theory, it is possible to show the existence of the adiabatic limit, see references in~\cite{froebcadamuro2022a}. However, the construction does not generalize easily to Lorentzian signature; see~\cite{bahnspinamontirejzner2023} for partial results in the massive theory.}. For de Sitter spacetime, where the spatial sections are compact, this limit only needs to be taken in the timelike direction, and it is possible that this simplifies the analysis.

\begin{acknowledgments}
This work has been funded by the Deutsche Forschungsgemeinschaft (DFG, German Research Foundation) --- project no. 396692871 within the Emmy Noether grant CA1850/1-1.
\end{acknowledgments}

\appendix

\section{Finite symmetry transformations}
\label{sec:appendix_ds}

Here, we derive the expressions for a finite de~Sitter symmetry transformation, of the form~\eqref{eq:desitter_finite_transformation}
\begin{equation}
\label{eq:appendix_ds_finite_transformation}
\mathe^{a \xi_\text{rot} + b \xi_\text{boost,1} + c \xi_\text{boost,2}} f(\tau,\theta) = f(\tau_{abc}, \theta_{abc}) \eqend{.}
\end{equation}
This is most easily done by passing to the embedding space, performing the transformation, and restricting the result to the de~Sitter hyperboloid~\eqref{eq:desitter_hyperboloid}. We thus use the inverse embedding~\eqref{eq:desitter_embedding_inversion} to promote $f$ to a function of the $X^A$, and use that $\xi_\text{rot}$ is the restriction of $M^{12}$ and that $\xi_\text{boost,i}$ are the restriction of $M^{0i}$ to the hyperboloid. Equation~\eqref{eq:appendix_ds_finite_transformation} thus reads
\begin{equation}
\mathe^{a M^{12} + b M^{01} + c M^{02}} f(X) = f(X_{abc}) \eqend{,}
\end{equation}
and using that $M^{AB} = X^A \partial_{X_B} - X^B \partial_{X_A}$, one verifies that the solution is given by
\begin{equations}
X^0_{abc} &= \frac{a^2 - ( b^2 + c^2 ) \cos s}{s^2} X^0 + \frac{a c ( 1 - \cos s ) + b s \sin s}{s^2} X^1 + \frac{a b ( \cos s - 1 ) + c s \sin s}{s^2} X^2 \eqend{,} \\
X^1_{abc} &= \frac{a c ( \cos s - 1 ) + b s \sin s}{s^2} X^0 + \frac{ ( a^2 - b^2 ) \cos s - c^2}{s^2} X^1 + \frac{b c ( 1 - \cos s ) - a s \sin s}{s^2} X^2 \eqend{,} \\
X^2_{abc} &= \frac{a b ( 1 - \cos s ) + c s \sin s}{s^2} X^0 + \frac{b c ( 1 - \cos s ) + a s \sin s}{s^2} X^1 + \frac{( a^2 - c^2 ) \cos s - b^2}{s^2} X^2 \eqend{,}
\end{equations}
where we defined
\begin{equation}
s^2 \equiv a^2 - b^2 - c^2 \eqend{.}
\end{equation}
Finally, combining the embedding~\eqref{eq:desitter_embedding} and the inverse embedding~\eqref{eq:desitter_embedding_inversion} with these results, we obtain the transformed de~Sitter coordinates
\begin{equations}[eq:appendix_ds_coordinates]
\begin{split}
\tau_{abc} &= \arccot\Bigg[ \frac{\left[ a^2 - ( b^2 + c^2 ) \cos s \right]}{s^2} \cot \tau \\
&\qquad\qquad- \frac{\left[ a c ( 1 - \cos s ) + b s \sin s \right] \cos \theta + \left[ a b ( \cos s - 1 ) + c s \sin s \right] \sin \theta}{s^2 \sin \tau} \Bigg] \eqend{,}
\end{split} \\
\theta_{abc} &= \arctan\bigg[ \frac{- \left[ a b ( 1 - \cos s ) + c s \sin s \right] \cos \tau + \left[ b c ( 1 - \cos s ) + a s \sin s \right] \cos \theta + \left[ ( a^2 - c^2 ) \cos s - b^2 \right] \sin \theta}{- \left[ a c ( \cos s - 1 ) + b s \sin s \right] \cos \tau + \left[ ( a^2 - b^2 ) \cos s - c^2 \right] \cos \theta + \left[ b c ( 1 - \cos s ) - a s \sin s \right] \sin \theta} \bigg] \eqend{.}
\end{equations}
These expressions are real for both $s^2 > 0$ (when the rotations dominates over the boosts), and for $s^2 < 0$ (when the boosts dominate over the rotation). In the latter case, this is seen by writing $s = \mathi t$ with $t \in \mathbb{R}$, and using that $\cos s = \cosh t$ and $s \sin s = - t \sinh t$.

For small transformations $\abs{a}, \abs{b}, \abs{c} \ll 1$, we obtain to first order
\begin{equations}[eq:appendix_ds_coordinates_leading]
\tau_{abc} &\approx \tau + b \cos \theta \sin \tau + c \sin \theta \sin \tau \eqend{,} \\
\theta_{abc} &\approx \theta + a + b \sin \theta \cos \tau - c \cos \theta \cos \tau \eqend{,}
\end{equations}
which is of course nothing else but
\begin{equations}
\tau_{abc} &\approx \tau + \left[ a \xi_\text{rot} + b \xi_\text{boost,1} + c \xi_\text{boost,2} \right] \tau \eqend{,} \\
\theta_{abc} &\approx \theta + \left[ a \xi_\text{rot} + b \xi_\text{boost,1} + c \xi_\text{boost,2} \right] \theta \eqend{.}
\end{equations}

Finally, one verifies in a long but straightforward computation that
\begin{equation}
\label{eq:appendix_ds_finite_transformation_inv}
\mathe^{- a \xi_\text{rot} - b \xi_\text{boost,1} - c \xi_\text{boost,2}} f(\tau_{abc},\theta_{abc}) = f(\tau, \theta) \eqend{,}
\end{equation}
such that the inverse transformation is obtained by inverting the parameters, and that
\begin{equation}
\label{eq:appendix_ds_measure}
\sqrt{-g} \total^2 x = \frac{1}{H^2 \sin^2(\tau)} \total \tau \total \theta = \frac{1}{H^2 \sin^2(\tau_{abc})} \total \tau_{abc} \total \theta_{abc} = \sqrt{-g(x_{abc})} \total^2 x_{abc} \eqend{,}
\end{equation}
as required since rotations and boosts are symmetry transformations.

\section{Cosine bound}
\label{sec:appendix_cosine}

Here, we want to derive the bound
\begin{equation}
\label{eq:appendix_cosine_bound}
\frac{1}{2 - 2 \cos(x)} \leq \begin{cases} \frac{\pi^2}{8 x^2} & x \in \left[ - \frac{\pi}{2}, \frac{\pi}{2} \right] \\ \frac{1}{2} & x \in \left[ \frac{\pi}{2}, \frac{3 \pi}{2} \right] \eqend{.} \end{cases}
\end{equation}
Since $\cos(x) < 0$ for $x \in [ \pi/2, 3\pi/2 ]$, the second part is obvious. The first part follows from the results of~\cite{andersonetal2006,pinelis2001} (see~\cite{yang2014} for an overview), and was explicitly derived in~\cite{bagulpanchal2022}; however, we provide a more direct proof.

Since the cosine is an even function, it is enough to show the bound for $x \in [0,\pi/2]$. We use the known facts that $\sin x$ and $\cos x$ are non-negative and bounded by $1$ and that $\sin x$ is monotonously increasing and $\cos x$ monotonously decreasing on that interval. Then we have that $x \sin x \geq 0$ and thus its integral $\int_0^x t \sin t \total t = \sin x - x \cos x$ is monotonously increasing, vanishes at $x = 0$, and is thus non-negative. Therefore, also the function $g(x) = 2 (\sin x - x \cos x) (1-\cos x)/\sin^2 x$ is non-negative since all factors are, and its integral $\int_0^x g(t) \total t = 2 x (1-\cos x)/\sin x - x^2$ is monotonously increasing, vanishes at $x = 0$, and is thus non-negative. Now we note that
\begin{equation}
\left( \frac{x^2}{1-\cos x} \right)' = \frac{\sin x}{(1-\cos x)^2} \left[ \frac{2 x (1-\cos x)}{\sin x} - x^2 \right] \geq 0 \eqend{,}
\end{equation}
so the function $x^2/(1-\cos x)$ is monotonously increasing. It is thus bounded by its value at $x = \pi/2$, which gives the bound~\eqref{eq:appendix_cosine_bound}.

\subsection*{Conflict of interest statement}

On behalf of all authors, the corresponding author states that there is no conflict of interest.

\subsection*{Data availability statement}

This manuscript has no associated data.

\bibliography{literature}

\end{document}